\documentclass[MSNbibl,nameyear,seceqn,dvips]{arxstspdf}
\usepackage{graphicx}
\usepackage{flushend}
\usepackage{stfloats}
\usepackage{url,breakurl}
\volume{29}
\issue{2}
\pubyear{2014}
\firstpage{267}
\lastpage{284}
\doi{10.1214/13-STS454} %kopijuoti is PTS
\makeatletter
\renewcommand{\citep}[1]{\citeauthor{#1}, \citeyear{#1}}
\makeatother

\begin{document}
\begin{frontmatter}

\title{Comments on the Neyman--Fisher Controversy and Its Consequences}%
% kai straipsnis turi susijusiu diskusiju ir rejoinder'iu
%rejoinder at \relateddoi{r}{10.1214/00-STSXXXX}.}
\runtitle{The Neyman--Fisher Controversy}

\begin{aug}
\author[a]{\fnms{Arman} \snm{Sabbaghi}\corref{}\ead[label=e1]{armansabbaghi.stat@gmail.com}}
\and
\author[a]{\fnms{Donald B.} \snm{Rubin}\ead[label=e2]{rubin@stat.harvard.edu}}

\affiliation{Harvard University}

\runauthor{A. Sabbaghi and D.~B. Rubin}

\address[a]{Arman Sabbaghi is Assistant Professor of Statistics, Department of Statistics, Purdue
University, 250 N. University Street, West Lafayette, Indiana 47907,
USA \printead{e1}.
Donald B. Rubin is John L. Loeb Professor of Statistics, Department of Statistics,
Harvard University,
1 Oxford Street Fl. 7,
Cambridge, Massachusetts 02138, USA
\printead{e2}.}
\end{aug}

% ABSTRACT
%
\begin{abstract}
The Neyman--Fisher controversy considered here originated with the 1935
presentation of Jerzy Neyman's
\textit{Statistical Problems in Agricultural Experimentation} to the
Royal Statistical Society. Neyman asserted that the
standard ANOVA F-test for randomized complete block designs is valid,
whereas the analogous test for Latin squares is
invalid in the sense of detecting differentiation among the treatments,
when none existed on average, more often than
desired (i.e., having a higher Type I error than advertised). However,
Neyman's expressions for the expected mean residual
sum of squares, for both designs, are generally incorrect. Furthermore,
Neyman's belief that the Type I error (when testing
the null hypothesis of zero average treatment effects) is higher than
desired, whenever the expected mean treatment sum of
squares is greater than the expected mean residual sum of squares, is
generally incorrect. Simple examples show that,
without further assumptions on the potential outcomes, one cannot
determine the Type I error of the F-test from expected
sums of squares. Ultimately, we believe that the Neyman--Fisher
controversy had a deleterious impact on the development of
statistics, with a major consequence being that potential outcomes were
ignored in favor of linear models and classical
statistical procedures that are imprecise without applied contexts.
\end{abstract}

% KEYWORDS
% Pirmas kwd is didziosios raides
%

\end{frontmatter}

%s1 #&#
%s1 ###
\section{Conflict and Controversy}
\label{sec:intro}

Prior to the presentation of \textit{Statistical Problems in
Agricultural Experimentation} to the Royal
Statistical Society in $1935$ (\citep{Neyman1935}), Jerzy Neyman and
Ronald Aylmer Fisher were on fairly good terms, both
professionally and personally. Joan Fisher Box's biography of her
father
(\citep{Box}, pages 262--263, 451) and Neyman's oral
autobiography (\citep{Reid}, pages 102, 114--117) describe two
scientists who respected each other during this time. However,
Neyman's study of randomized complete block (RCB) and Latin square (LS)
designs sparked Fisher's legendary temper
(\citep{Reid}, pages 121--124; \citep{Box}, pages 262--266;
\citep{Lehmann}, pages 58--59), with the resulting heated debate
recorded in the discussion. The relationship between Fisher and Neyman
became acrimonious, with no reconciliation ever
being reached (\citep{Reid}, pages 124--128, 143, 183--184, 225--226, 257;
\citep{Lehmann}, Chapter~4).

The source of this conflict was Neyman's suggestion that RCBs were a
more valid experimental design than LSs, for both
hypothesis testing and precision of estimates. He reached this
conclusion using potential outcomes, which he introduced in
$1923$ as part of his doctoral dissertation (\citep{Neyman1923}), the
first place formalizing, explicitly, the notation of
potential outcomes for completely randomized (CR) experiments. \citet
{Neyman1935} extended this framework in a natural way
from CR designs to RCBs and LSs, and calculated the expected mean
residual sum of squares and expected mean treatment
sum of squares for both.

\citet{Neyman1935} stated that, under the null hypothesis of zero
average treatment effects (\emph{Neyman's null
hypothesis}), the expected mean residual sum of squares equals the
expected mean treatment sum of squares for RCBs, whereas
the expected mean residual sum of squares is less than or equal to the
expected mean treatment sum of squares for LSs, with
equality holding under special cases, such as \emph{Fisher's sharp null
hypothesis} of no individual treatment effects.
From this comparison of the expected mean residual and treatment sums
of squares, Neyman concluded that the standard ANOVA
F-test for RCBs was ``unbiased,'' whereas the corresponding test for
LSs was ``biased,'' potentially detecting
differentiation among the treatments, when none existed on average,
more often than desired (i.e., having a higher Type I
error than advertised under Neyman's null):

\begin{quote}
In the case of the Randomized Blocks the position is somewhat more
favourable to the z test [i.e., the
F-test], while in the case of the Latin Square this test seems to be
biased, showing the tendency to discover
differentiation when it does not exist. It is probable that the
disturbances mentioned are not important from the point of
view of practical applications. (\citep{Neyman1935}, page 114)
\end{quote}

Fisher's fury at Neyman's assertions is evident in his transcribed response:

\begin{quote}
Professor R. A. Fisher, in opening the discussion, said he had hoped
that Dr. Neyman's paper would be on a
subject with which the author was fully acquainted, and on which he
could speak with authority \ldots. Since seeing
the paper, he had come to the conclusion that Dr. Neyman had been
somewhat unwise in his choice of topics.
\ldots
Apart from its theoretical defects, Dr. Neyman appears also to have
discovered that it [the LS]
was, contrary to general belief, a less precise method of
experimentation than was supplied by Randomized Blocks,
even in those cases in which it had hitherto been regarded as the more
precise design. It appeared, too, that they
had to thank him, not only for bringing these discoveries to their
notice, but also for concealing them from public
knowledge until such time as the method should be widely adopted in practice!
\ldots
I think it is clear to everyone present that Dr. Neyman has
misunderstood the intention \ldots of the z test
and of the Latin Square and other techniques designed to be used with
that test. Dr. Neyman thinks that another test
would be more important. I am not going to argue that point. It may be
that the question which Dr. Neyman thinks
should be answered is more important than the one I have proposed and
attempted to answer. I suggest that before
criticizing previous work it is always wise to give enough study to the
subject to understand its purpose. Failing
that it is surely quite unusual to claim to understand the purpose of
previous work better than its author.
(\citep{Fisher1935}, pages 154, 155, 173)
\end{quote}

Although Fisher reacted in an intemperate manner, his discussion
nevertheless hints at errors in Neyman's calculations. In
fact, Fisher was the sole discussant who identified an incorrect
equation (27), in Neyman's appendix:

\begin{quote}
Then how had Dr. Neyman been led by his symbolism to deceive himself on
so simple a question? \ldots
Equations (13) and (27) of his appendix showed that the quantity
which Dr. Neyman had chosen to call $\sigma^2$ did not
contain the same components of error as those which affected the actual
treatment means, or as those which contributed to
the estimate of error. (\citep{Fisher1935}, page 156)
\end{quote}

Neyman in fact made a crucial algebraic mistake in his appendix, and
his expressions for the expected mean residual sum of
squares for both designs are generally incorrect. We present the
correct expressions in Sections~\ref{sec:RCBD_theory} and
\ref{sec:LSD_theory}, and provide an interpretation of these formulae
in Section~\ref{sec:interactions_EMS}. As we shall
see, if one subscribes to Neyman's suggestion that a comparison of
expected mean sums of squares determines Type I errors
when testing Neyman's null, then the F-test for RCBs is predictably
wrong, whereas the F-test for LSs is unpredictably
wrong.

However, Neyman's suggestion is generally incorrect. We present in
Section~\ref{sec:concrete} simple
examples of LSs for which Neyman's null holds and the expected mean
residual sum of squares equals the expected mean
treatment sum of squares, yet the Type I error of the F-test is smaller
than nominal. Such examples lead to the general
result that, for any size RCB or LS, Type I errors are not dictated by
a simple comparison of expected sums of squares
without further conditions.

A cacophony of commentary on this controversy exists in the literature,
and we compiled the most relevant articles in
Sections~\ref{sec:RCBD_history}, \ref{sec:LSD_history} and \ref
{sec:cacophony}. Our results agree with similar
calculations made by \citet{Wilk1955} and \citet{WilkKempthorne}. A
major difference is that we work in a more general
setting of Neyman's framework, whereas others [especially \citet
{Wilk1955}] tend to make further assumptions on
the potential outcomes, albeit assumptions possibly justified by
applied considerations. Furthermore, although
\citet{WilkKempthorne} extend Neyman's framework to consider random
sampling of rows, columns and treatment levels from
some larger population for LSs, their ultimate suggestion that the
expected mean residual sum of squares is larger
than the expected mean treatment sum of squares is not generally true.
A different parametrization of similar quantities,
used in Section~\ref{sec:interactions_EMS}, reveals how the inequality
could go in either direction.

This controversy had substantial consequences for the subsequent
development of statistics for experimental design. As we
discuss in Section~\ref{sec:consequences}, deep issues arising from
this disagreement led to a shift from \emph{potential
outcomes} to additive models for \emph{observed outcomes} in
experiments, seriously limiting the scope of inferential tools
and reasoning. Our ultimate goal in this historical study is not simply
to correct Neyman's algebra. Instead, we wish to
highlight the genesis of the current approach to experimental design
resulting from this controversy, which is based on
linear models and other simple regularity conditions on the potential
outcomes that are imprecise without applied contexts.

%s2 #&#
%s2 ###
\section{Controversial Calculations}
\label{sec:ANOVA}

%s2.1 #&#
%s2.1 ###
\subsection{Randomized Complete Block Designs: Theory}
\label{sec:RCBD_theory}

We first consider RCBs with $N$ blocks, indexed by $i$, and $T$
treatments, indexed by $t$, with each block having $T$
experimental units, indexed by $j = 1, \ldots, T$. Treatments are
assigned randomly to units in a block, and are applied
independently across blocks (\citep{Kempthorne}, Chapter~9). Although
our results hold for general RCB designs, we adopt the
same context as Neyman: blocks represent physical blocks of land on a
certain field, and we compare agricultural treatments
that may affect crop yield, for example, fertilizers.

We explicitly define treatment indicators $\mathbf{W} =  \{W_{ij}(t)\}$ as
\begin{eqnarray*}
&&W_{ij}(t)\\
&& = \cases{1, & $\mbox{if unit } j \mbox{ in block } i
\mbox{ is assigned treatment } t,$ \vspace*{2pt}
\cr
0, & $\mbox{otherwise.}$ }
\end{eqnarray*}
\citet{Neyman1935} specified the potential outcomes as
\[
x_{ij}(t) = X_{ij}(t) + \varepsilon_{ij}(t),
\]
where $X_{ij}(t) \in\mathbb{R}$ are unknown constants representing the
``mean yield'' of unit $j$ in block $i$
under treatment~$t$, and $\varepsilon_{ij}(t) \sim[0, \sigma_{\varepsilon
}^2]$ are mutually independent and identically
distributed (i.i.d.) ``technical errors,'' independent of the random
variables \textbf{W}. This framework for the potential
outcomes, excluding the $\varepsilon_{ij}(t)$, is similar to that
presented in Neyman's 1923 dissertation
(\citep{Neyman1923}).

\citeauthor{Neyman1935} [(\citeyear{Neyman1935}), pages 110, 114, 145]
stated that technical errors
represent inaccuracies in the experimental technique,
for example, inaccuracies in measuring crop yield, and assumed that
technical errors are Normal random variables. We find these
technical errors rather obscure, but their inclusion does not alter our
conclusions. To summarize, in Neyman's
specification there are two sources of randomness: the unconfounded
assignment mechanism
(\citep{Rubin1990}), that is, the random
assignment of treatments to plots specified by the distribution on
$\mathbf{W}$, and the technical errors $\varepsilon_{ij}(t)$.

Potential outcomes are decomposed by \citeauthor{Neyman1935}
[(\citeyear{Neyman1935}), page 111] into
%
%e2.1 #&#
%e2.1 ###
\begin{equation}
x_{ij}(t) = \bar{X}_{\cdot\cdot}(t) + B_i(t) +
\eta_{ij}(t) + \varepsilon_{ij}(t),
\end{equation}
where
\begin{eqnarray*}
\bar{X}_{\cdot\cdot}(t)& = &\frac{1}{NT}\sum_{i=1}^N
\sum_{j=1}^T X_{ij}(t),
\\
B_i(t)& =& \bar{X}_{i \cdot}(t) - \bar{X}_{\cdot\cdot}(t),
\\
\eta_{ij}(t) &=& X_{ij}(t) - \bar{X}_{i \cdot}(t),
\end{eqnarray*}
with
\[
\bar{X}_{i \cdot}(t) = \frac{1}{T} \sum_{j=1}^T
X_{ij}(t).
\]
Neyman describes $B_i(t)$ as a correction for the specific fertility of
the $i$th block, and $\eta_{ij}(t)$ as
a correction for fertility variation within the block or,
alternatively, the soil error.
Hinkelmann and Kempthorne [(\citeyear{Kempthorne}), page 300] refer
to terms such as $\eta_{ij}(t)$ as unit-treatment interactions, but
they distinguish between \emph{strict} unit-treatment interactions and
block-treatment interactions. For strict unit-treatment interaction,
treatment effects depend on the
experimental unit, in the sense that for two treatments $t, t'$ and
experimental units $j, j'$ in a block $i$,
\[
X_{ij}(t) - X_{ij}\bigl(t'\bigr) \neq
X_{ij'}(t) - X_{ij'}\bigl(t'\bigr).
\]
Block-treatment interactions are characterized by treatment effects
depending\vadjust{\goodbreak} on the block, in the sense that for two
treatments $t,t'$, experimental units $j, j', j'',  j'''$, and blocks
$i, i'$,
\[
X_{ij}(t) - X_{ij'}\bigl(t'\bigr) \neq
X_{i'j''}(t) - X_{i'j''''}\bigl(t'\bigr).
\]
As pointed out by a referee, allowing fertility variation to depend on
treatment $t$ was a unique contribution by Neyman and
was never recognized in the discussion by Fisher, who focused on his
sharp null hypothesis (described next), under which the
corrections do not depend on $t$.

The purpose of the local field experiment, as described by
\citeauthor{Neyman1935} [(\citeyear{Neyman1935}), page 111] is to compare the
$\bar{X}_{\cdot\cdot}(t)$ for $t = 1, \ldots, T$, each of which
represents the average mean yield when one treatment $t$
is applied to all plots in the field, a conceptual experiment. As
stated in the discussion, and later by
\citeauthor{Welch} [(\citeyear{Welch}), page 23]
Neyman does not test \emph{Fisher's sharp null
hypothesis} of zero individual treatment effects,
that is (when excluding technical errors),
%
%e2.2 ###
\begin{eqnarray}
H_0^{\#}\dvtx X_{ij}(t) = X_{ij}
\bigl(t'\bigr) \nonumber\\
\eqntext{\forall i = 1, \ldots, N; j = 1, \ldots, T; t \neq
t'.}
\end{eqnarray}
Instead, Neyman sought to test the more general null hypothesis
\[
H_0\dvtx\bar{X}_{\cdot\cdot}(1) = \cdots = \bar{X}_{\cdot\cdot}(T),
\]
referred to throughout as \emph{Neyman's null hypothesis}:

\begin{quote}
I am considering problems which are important from the point of view of
agriculture. And from this viewpoint
it is immaterial whether any two varieties react a little differently
to the local differences in the soil. What is
important is whether on a larger field they are able to give equal or
different yields.
(\citep{Neyman1935}, page 173)
\end{quote}

\noindent If the treatment effects are additive across all units, that is,
%
%e2.3 ###
\begin{eqnarray}
X_{ij}(t) = U_{ij} + \tau(t)\nonumber\\
\eqntext{ \forall i = 1, \ldots, N; j =
1, \ldots, T; t = 1, \ldots, T,}
\end{eqnarray}
then testing Neyman's null is equivalent to testing Fisher's sharp null.

The observed yield of the plot assigned treatment $t$ in block $i$ is
\[
y_i(t) = \sum_{j=1}^T
W_{ij}(t) x_{ij}(t),
\]
and the observed average yield for all plots assigned treatment $t$ is
\[
\bar{y}_{\cdot}(t) = \frac{1}{N} \sum_{i=1}^N
y_i(t).\vadjust{\goodbreak}
\]
\citeauthor{Neyman1935} [(\citeyear{Neyman1935}), page 112]
noted that an unbiased estimator for the
difference between average treatment means,
$\bar{X}_{\cdot\cdot}(t) - \bar{X}_{\cdot\cdot}(t')$, is $\bar
{y}_{\cdot}(t) - \bar{y}_{\cdot}(t')$,
and correctly calculated its sampling variance over its randomization
distribution as
\begin{eqnarray*}
\operatorname{Var} \bigl\{ \bar{y}_{\cdot}(t) - \bar{y}_{\cdot}
\bigl(t'\bigr) \bigr\} &=& \frac{2\sigma_{\varepsilon}^2}{N}
 + \frac{\sigma_{\eta}^2(t) + \sigma_{\eta}^2(t')}{N}\\
 &&{} +
\frac{2r(t,t')\sqrt{\sigma_{\eta}^2(t) \sigma_{\eta}^2(t')}}{N(T-1)},
\end{eqnarray*}
where
\begin{eqnarray*}
\sigma_{\eta}^2(t) &=& \frac{1}{NT}\sum
_{i=1}^N \sum_{j=1}^T
\eta _{ij}(t)^2,
\\
r\bigl(t,t'\bigr)& =& \frac{\sum_{i=1}^N \sum_{j=1}^T \eta_{ij}(t) \eta
_{ij}(t')}{NT \sqrt{\sigma_{\eta}^2(t)
\sigma_{\eta}^2(t')}}.
\end{eqnarray*}
\citeauthor{Neyman1935} [(\citeyear{Neyman1935}), page 145] assumed
that $\sigma_{\eta}^2(t)$ and
$r(t,t')$ are constant functions of
$t, t'$ only to save space and simplify later expressions; this
particular set of assumptions appears to have been made
purely for mathematical simplicity, and is not driven by any applied
considerations, unlike assumptions made by
\citet{Wilk1955} and \citet{WilkKempthorne} (described in Sections~\ref{sec:RCBD_history} and \ref{sec:LSD_history}).

Neyman then calculated expectations of mean residual sum of squares and
mean treatment sum of squares, expressed in our
notation as (resp.)
\begin{eqnarray*}
S_0^2 &= &\frac{1}{(N-1)(T-1)}\\
&&{}\times \sum
_{i=1}^N \sum_{t=1}^T
\bigl\{ y_i(t) - \bar{y}_{\cdot}(t) - \bar
{y}_i(\cdot) + \bar{y}_{\cdot}(\cdot) \bigr\}^2
\end{eqnarray*}
and
\[
S_1^2 = \frac{N}{T-1}\sum
_{t=1}^T \bigl\{ \bar{y}_{\cdot}(t) -
\bar{y}_{\cdot
}(\cdot) \bigr\}^2.
\]
As proven in our appendix (\citep{SabbaghiRubin}), the expectations are
\begin{eqnarray*}
\mathbb{E}\bigl(S_0^2\bigr) &=& \sigma_{\varepsilon}^2
+ \frac{1}{T}\sum_{t=1}^T
\sigma_{\eta}^2(t) \\
&&{}+ \frac{1}{T(T-1)^2}\sum
_{t \neq t'}r\bigl(t,t'\bigr)\sqrt{
\sigma_{\eta}^2(t) \sigma_{\eta}^2
\bigl(t'\bigr)}
\\
&&{} + \frac{1}{(N-1)(T-1)} \sum_{i=1}^N \sum
_{t=1}^T \bigl\{ B_i(t) - \bar
{B}_i(\cdot) \bigr\}^2
\end{eqnarray*}
and
\begin{eqnarray*}
\mathbb{E}\bigl(S_1^2\bigr) &=& \sigma_{\varepsilon}^2
+ \frac{1}{T}\sum_{t=1}^T
\sigma_{\eta}^2(t)\\
&&{} + \frac{1}{T(T-1)^2}\sum
_{t \neq t'}r\bigl(t,t'\bigr) \sqrt{
\sigma_{\eta
}^2(t)\sigma_{\eta}^2
\bigl(t'\bigr)}
\\
&&{} + \frac{N}{T-1}\sum_{t=1}^T \bigl\{
\bar{X}_{\cdot\cdot}(t) - \bar{X}_{\cdot
\cdot}(\cdot) \bigr\}^2.
\end{eqnarray*}

\citeauthor{Neyman1935} [(\citeyear{Neyman1935}), pages 147--150]
correctly calculated the expected
mean treatment sum of squares, but made a mistake when
calculating the expected mean residual sum of squares. His incorrect
expression is equation (27) on page $148$.
\citeauthor{Sukhatme} [(\citeyear{Sukhatme}), page 166]
his Ph.D. student at the University of
London, incorrectly calculated the expectation for
the general case when $\sigma_{\eta}^2(t)$ and $r(t,t')$ are not
constant in $t,t'$, and the corresponding incorrect
expression is his equation (3):
\begin{eqnarray*}
&&\sigma_{\varepsilon}^2 + \frac{1}{T}\sum
_{t=1}^T \sigma_{\eta}^2(t)
\\
&&\quad{}+\frac{1}{T(T-1)^2}\sum_{t \neq t'}r\bigl(t,t'
\bigr)\sqrt{\sigma_{\eta}^2(t) \sigma_{\eta}
\bigl(t'\bigr)}.
\end{eqnarray*}

To see why the last term in $\mathbb{E}(S_0^2)$ is missing in these
equations, note that the expression within the
brackets of $S_0^2$ can be written as the sum of the three terms
\begin{eqnarray*}
&&B_i(t) - \bar{B}_i(\cdot),
\\
&&\sum_{j=1}^T W_{ij}(t)
\eta_{ij}(t)  - \frac{1}{N} \sum_{i=1}^N
\sum_{j=1}^T W_{ij}(t)
\eta_{ij}(t) \\
&&\quad{}- \frac{1}{T}\sum_{t=1}^T
\sum_{j=1}^T W_{ij}(t)
\eta_{ij}(t)
\\
&&\quad{} + \frac{1}{NT} \sum_{i=1}^N \sum
_{t=1}^T \sum_{j=1}^T
W_{ij}(t)\eta_{ij}(t)
\end{eqnarray*}
and
\begin{eqnarray*}
&&\sum_{j=1}^T W_{ij}(t)
\varepsilon_{ij}(t)  - \frac{1}{N} \sum
_{i=1}^N \sum_{j=1}^T
W_{ij}(t) \varepsilon_{ij}(t)\\
&&\quad{} - \frac{1}{T}\sum
_{t=1}^T \sum_{j=1}^T
W_{ij}(t) \varepsilon_{ij}(t)
\\
&&\quad{} + \frac{1}{NT} \sum_{i=1}^N \sum
_{t=1}^T \sum_{j=1}^T
W_{ij}(t) \varepsilon_{ij}(t).
\end{eqnarray*}
Neyman's equation $(17)$ is missing the first term $B_i(t) - \bar
{B}_i(\cdot)$, which is not necessarily equal to
zero, and was never explicitly declared to be zero by Neyman.

Consequently, under Neyman's null, the expected mean residual sum of
squares is greater than or equal to the expected mean
treatment sum of squares, with equality holding if and only if for each
block $i$, $B_i(t)$ is constant across treatments
$t$. Alternatively, equality holds under Fisher's sharp null. If one
accepts Neyman's logic regarding ``unbiased tests''
(discussed in Section~\ref{sec:cacophony}), then the correct
expressions for the expectations of mean squares suggest that
the standard ANOVA F-test for RCBs has a Type I error bounded above by
its nominal level.

A simple example makes this concrete. Suppose $N = T = 2$ and $\sigma
_{\varepsilon}^2 = 0$, with the potential outcomes in
Table~\ref{tab1}. Note that $\bar{X}_{\cdot\cdot}(\mathrm{1}) = \bar
{X}_{\cdot\cdot}(\mathrm{2})$, so Neyman's null is
satisfied. We calculate $\mathbb{E}(S_0^2) = 215.875, \mathbb{E}(S_1^2)
= 213.625$, and
\[
\mathbb{E}\bigl(S_0^2\bigr) - \mathbb{E}
\bigl(S_1^2\bigr) = 2.25 = \sum
_{i=1}^2\sum_{t=1}^2
\bigl\{ B_i(t) - \bar{B}_i(\cdot) \bigr
\}^2.
\]

%t1 #&#
%t1 ###
\begin{table}
\caption{Table of potential outcomes for a RCB with $\mathbb{E}(S_0^2)
> \mathbb{E}(S_1^2)$}\label{tab1}
\begin{tabular*}{\columnwidth}{@{\extracolsep{\fill}}lcc@{}}
\hline
& \textbf{Treatment} $\bolds{1}$ & \multicolumn{1}{c@{}}{\textbf{Treatment} $\bolds{2}$} \\
\hline
Block $1$, Plot $1$ & $10$ & $15$ \\
Block $1$, Plot $2$ & $10$ & $2$ \\
Block $2$, Plot $1$ & $20$ & $3$ \\
Block $2$, Plot $2$ & $30$ & $50$ \\
\hline
\end{tabular*}
\end{table}

%s2.2 #&#
%s2.2 ###
\subsection{Randomized Complete Block Designs: After the Controversy}
\label{sec:RCBD_history}

Neyman's potential outcomes framework is similar to the ``conceptual
yield'' framework developed by
\citeauthor{Kempthorne1952} (\citeyear{Kempthorne1952,Kempthorne1955}).
Certain features of these two
are only cosmetically different: for example,
\citeauthor{Kempthorne1952} [(\citeyear{Kempthorne1952}), page 137] and
later\vadjust{\goodbreak} \citeauthor{Kempthorne} [(\citeyear{Kempthorne}), page
280] represent treatment indicators by $\delta_{ij}^k$
(with $k$ denoting treatment level) and potential outcomes as
$y_{ijk}$. As emphasized by a referee, using treatment
indicators as random variables provides a mathematical foundation for
the randomization theory of \citet{Fisher}, connecting
potential outcomes with observed responses.

An important difference between Neyman and Kempthorne concerns the
notion of technical errors.
\citeauthor{Kempthorne} [(\citeyear{Kempthorne}), page 161]
make a distinction between experimental and observational errors, and
include separate terms for each, allowing them to
depend on treatment. Neyman effectively only considers their sum when
defining technical errors, which may be a source of
confusion. Of course, Neyman's results were for local field
experiments, in which case he might not have considered it
necessary to introduce observational errors arising from random
sampling of experimental units from some larger population.

\citet{Kempthorne1952} made an interesting comment relating to Fisher's
sharp null, Neyman's null and Neyman's notation
for technical errors:

\begin{quote}
If the experimenter is interested in the more fundamental research
work, Fisher's null hypothesis is more satisfactory, for
one should be interested in discovering the fact that treatments have
different effects on different plots and in trying to
explain why such differences exist. It is only in technological
experiments designed to answer specific questions about a
particular batch of materials which is later to be used for production
of some sort that Neyman's null hypothesis appears
satisfactory \ldots Neyman's hypothesis appears artificial in this
respect, that a series of repetitions is envisaged, the
experimental conditions remaining the same but the technical errors
being different.
(\citep{Kempthorne1952}, page 133)
\end{quote}

Furthermore, \citeauthor{Kempthorne1952} [(\citeyear{Kempthorne1952}),
pages 145--151] correctly noted (in
agreement with our results in Section~\ref{sec:interactions_EMS}) that block-treatment interactions must be
zero in order for
$\mathbb{E}(S_0^2) = \mathbb{E}(S_1^2)$ under Neyman's null, also known
as unbiasedness of a design in the \citet{Yates}
sense. As Kempthorne stated in a later article:

\begin{quote}
For the case of randomized blocks it is found that block treatment
interactions must be zero in order that the design be
unbiased in Yates's sense.\vadjust{\goodbreak} \ldots It does not appear to be at all
desirable to section the experimental material into ordinary
randomized blocks, of \ldots highly different fertilities (or basal
yields) because this procedure is likely to lead to block
treatment interactions. (\citep{Kempthorne1955}, page 964)
\end{quote}

Additivity of treatment effects was not invoked by Neyman, and
nonadditivity for RCBs was investigated later
(\citep{Tukey}; \citep{Kempthorne1955};
\citep{Wilk1955}; \citep{Mandel}). Perhaps the most substantial work,
in the same direction as Neyman, was
done by \citet{Wilk1955}, who extended the results of
\citeauthor{Kempthorne1952} [(\citeyear{Kempthorne1952}), pages~145--151] for RCBs to the case of
generalized randomized blocks. Wilk studied randomization moments of
mean sums of squares, estimation of various
finite-population estimands and Normal theory approximations for
testing Fisher's sharp null and Neyman's null. He also
distinguished between experimental error, that is, the failure of
different experimental units treated alike to respond
identically, and technical error, or limitations on experimental
technique that prevent the exact reproduction of an
applied treatment. To us, this use of notation confuses mathematical
derivations and practical interpretations of symbols.

More importantly, although Wilk made assumptions on the potential
outcomes (consequently not working in our more general
setting), he attempted to justify them as physically relevant, as
opposed to Neyman, who only made
assumptions to facilitate calculations. For example, when translating
Wilk's notation into Neyman's, we see that
\citeauthor{Wilk1955} [(\citeyear{Wilk1955}), page 72] explicitly considered the physical situation
that, if the blocking of experimental units is
successful, then the $\eta_{ij}(t) - \bar{\eta}_{ij}(\cdot)$ will be
negligible for all $i, j, t$, whereas block-treatment
interactions $B_i(t) - \bar{B}_i(\cdot)$ would be important, in the
sense of varying with $t$. When units in a block
are as homogeneous as possible with respect to background covariates,
the assumption of no strict unit-treatment interactions
becomes more plausible, similar to the plausibility of zero partial
correlation among potential outcomes given all measured
covariates. Accordingly, block-treatment interactions become more
important. A referee made a similar comment, remarking that
for agronomic experiments, it is reasonable to assume that the $\eta
_{ij}(t)$ are negligible, whereas in situations such as
medical experiments involving human subjects, this may no longer be true.

Wilk's explicit physical consideration is used to justify his
assumption (stated without further explanation by
\citeauthor{Kempthorne} [(\citeyear{Kempthorne}), page 301] in their description of the general model
for RCBs) that treatments react additively within a
block but\vadjust{\goodbreak} can react nonadditively from block-to-block, that is,
\begin{eqnarray*}
\bigl\{ X_{ij}(t) - \bar{X}_{ij}(\cdot) \bigr\} - \bigl\{
\bar{X}_{i \cdot}(t) - \bar {X}_{i \cdot}(\cdot) \bigr\} &=&
\eta_{ij}(t) - \bar{\eta}_{ij}(\cdot) \\
&=& 0
\end{eqnarray*}
for all $i, j, t$, even though
\[
B_i(t) - \bar{B}_i(\cdot) \neq0
\]
for at least one pair $(i, t)$. \citeauthor{Wilk1955}
[(\citeyear{Wilk1955}), page 73] then stated
that, if
\[
\eta_{ij}(t) - \bar{\eta}_{ij}(\cdot) \neq0
\]
for at least one triple $(i,j,t)$, then the expected mean treatment sum
of squares is not equal to the expected mean
residual sum of squares under Neyman's null.
\citeauthor{Kempthorne} [(\citeyear{Kempthorne}), page 301] when summarizing Wilk's work, noted that the
expected mean residual sum of squares for RCB designs contains the
interaction between blocking and treatment factors,
similar to our result.

%s2.3 #&#
%s2.3 ###
\subsection{Latin Square Designs: Theory}
\label{sec:LSD_theory}

It was in his treatment of LSs that Neyman's error substantially
changes conclusions. We consider $T \times T$ LSs with
rows and columns denoting levels of two blocking factors, for example,
north--south and east--west. Our treatment indicators are
\[
W_{ij}(t) = \cases{1, & $\mbox{if the unit in row } i,
\mbox{column } j,$\vspace*{2pt}\cr
&$\mbox{is assigned treatment } t$, \vspace*{2pt}
\cr
0, & $\mbox{otherwise.}$}
\]

Neyman specified the potential outcomes as
\[
x_{ij}(t) = X_{ij}(t) + \varepsilon_{ij}(t),
\]
with $X_{ij}(t) \in\mathbb{R}$ unknown constants representing the
``mean yield'' of the unit in cell $(i,j)$ under
treatment $t$, and $\varepsilon_{ij}(t) \sim[0, \sigma_{\varepsilon}^2]$
technical errors that are i.i.d. and independent of
\textbf{W}. Potential outcomes were then decomposed into
%
%e2.2 #&#
%e2.4 ###
\begin{eqnarray}
x_{ij}(t) &=& \bar{X}_{\cdot\cdot}(t) + R_i(t) +
C_j(t)
\nonumber
\\[-8pt]
\\[-8pt]
\nonumber
&&{}+ \eta_{ij}(t) + \varepsilon_{ij}(t),
\end{eqnarray}
where
\begin{eqnarray*}
R_i(t) &= &\bar{X}_{i \cdot}(t) - \bar{X}_{\cdot\cdot}(t),
\\
C_j(t) &=& \bar{X}_{\cdot j}(t) - \bar{X}_{\cdot\cdot}(t),
\\
\eta_{ij}(t) &=& X_{ij}(t) - \bar{X}_{i \cdot}(t) -
\bar{X}_{\cdot j}(t) + \bar{X}_{\cdot\cdot}(t).
\end{eqnarray*}
Similar to RCBs, Neyman described $R_i(t)$ and $C_j(t)$ as corrections
for specific soil
fertility of the $i$th row and $j$th column, respectively, and
$\eta_{ij}(t)$ as the soil error for plot
$(i,j)$ under treatment $t$.\vadjust{\goodbreak}

We define $\bar{x}_{\cdot\cdot}^o(t)$ as the observed average yield
for plots assigned treatment $t$,
\[
\bar{x}_{\cdot\cdot}^{o}(t) = \frac{1}{T} \sum
_{i=1}^T \sum_{j=1}^T
W_{ij}(t) x_{ij}(t).
\]
\citet{Neyman1935} correctly noted that
$\mathbb{E} \{ \bar{x}_{\cdot\cdot}^{o}(t) - \bar{x}_{\cdot\cdot
}^{o}(t') \} =
\bar{X}_{\cdot\cdot}(t) -  \bar{X}_{\cdot\cdot}(t')$ and that
\begin{eqnarray*}
&&\operatorname{Var}\bigl\{ \bar{x}_{\cdot\cdot}^{o}(t) -
\bar{x}_{\cdot\cdot
}^{o}\bigl(t'\bigr) \bigr\}\\
&&\quad=
\frac{2\sigma_{\varepsilon}^2}{T} +
\frac{\sigma_{\eta}^2(t) + \sigma_{\eta}^2(t')}{T-1}\\
&&\qquad{} + \frac{2r(t,t')\sqrt{\sigma_{\eta}^2(t) \sigma_{\eta}^2(t')}}{(T-1)^2}.
\end{eqnarray*}
Neyman then calculated the expected mean sums of squares. The mean
residual and treatment sums of squares are defined as
(resp.)
\begin{eqnarray*}
S_0^2 &=& \frac{1}{(T-1)(T-2)}\\
&&{}\times \sum
_{i=1}^T \sum_{j=1}^T
\Biggl\{ y_{ij} - \bar{y}_{i \cdot} - \bar {y}_{\cdot j} \\
&&\hspace*{50pt}{}-
\sum_{t=1}^T W_{ij}(t)
\bar{x}_{\cdot\cdot}^o(t) + 2 \bar{y}_{\cdot
\cdot} \Biggr
\}^2
\end{eqnarray*}
and
\[
S_1^2 = \frac{T}{T-1}\sum
_{t=1}^T \bigl\{ \bar{x}_{\cdot\cdot}^o(t)
- \bar {y}_{\cdot\cdot} \bigr\}^2,
\]
with $y_{ij} = \sum_{t=1}^T W_{ij}(t) x_{ij}(t)$ the observed response
of cell $(i,j)$, and
\begin{eqnarray*}
\bar{y}_{i \cdot} &=& \frac{1}{T} \sum_{j=1}^T
y_{ij},\\
 \bar{y}_{\cdot j}& =& \frac{1}{T} \sum
_{i=1}^T y_{ij}, \\
\bar{y}_{\cdot\cdot} &=&
\frac{1}{T} \sum_{j=1}^T
\bar{y}_{\cdot j} = \frac{1}{T} \sum_{i=1}^T
\bar{y}_{i
\cdot}
\end{eqnarray*}
We prove in our appendix (\citep{SabbaghiRubin}) that the correct
expectations are
\begin{eqnarray*}
\mathbb{E}\bigl(S_0^2\bigr) &=& \sigma_{\varepsilon}^2
+ \frac{T-2}{(T-1)^2}\sum_{t=1}^T
\sigma_{\eta}^2(t) \\
&&{}+ \frac{2}{(T-1)^3} \sum
_{t \neq t'}r\bigl(t,t'\bigr) \sqrt{
\sigma_{\eta}^2(t) \sigma_{\eta}^2
\bigl(t'\bigr)} \\
& &{} + \frac{1}{T(T-1)^2}\sum
_{i=1}^T \sum_{j=1}^T
\sum_{t=1}^T\bigl[\bigl\{
R_i(t) - \bar{R}_i(\cdot) \bigr\}^2 \\
&&\hspace*{110pt}{}+ \bigl
\{ C_j(t) - \bar{C}_j(\cdot) \bigr\}^2
\bigr]
\end{eqnarray*}
and
\begin{eqnarray*}
\mathbb{E}\bigl(S_1^2\bigr) &=& \sigma_{\varepsilon}^2
+ \frac{1}{T-1}\sum_{t=1}^T
\sigma_{\eta}^2(t) \\
&&{}+ \frac{1}{(T-1)^3} \sum
_{t \neq t'} r\bigl(t,t'\bigr)\sqrt{
\sigma_{\eta}^2(t) \sigma_{\eta}^2
\bigl(t'\bigr)}
\\
& &{}+ \frac{T}{T-1}\sum_{t=1}^T \bigl
\{ \bar{X}_{\cdot\cdot}(t) - \bar{X}_{\cdot
\cdot}(\cdot) \bigr
\}^2.
\end{eqnarray*}

\citeauthor{Neyman1935} [(\citeyear{Neyman1935}), page 152] made a similar mistake as he did for
RCBs, excluding
\[
R_i(t) + C_j(t) - \bar{R}_i(\cdot) -
\bar{C}_j(\cdot)
\]
in
a simplified expression\vspace*{1pt} for the term inside the brackets of $S_0^2$ in
his equation $(50)$. In effect, Neyman once
again excluded corrections for soil fertility, as it is not necessarily
true (nor stated explicitly) that $R_i(t)$ is
constant in $t$ for all rows $i$ and that $C_j(t)$ is constant in $t$
for all columns $j$.
\citeauthor{Sukhatme} [(\citeyear{Sukhatme}), page 167] made a
similar mistake for the case when $\sigma_{\eta}^2(t)$ and $r(t,t')$
are not constant in $t, t'$.

After incorrectly calculating the expected mean residual sum of
squares, Neyman stated that the expected mean residual sum
of squares was less than or equal to the expected mean treatment sum of
squares under Neyman's null
(\citep{Neyman1935}, page 154), with equality only under special cases,
such as Fisher's sharp null. Based on this observation,
Neyman conjectured that the standard ANOVA F-test for LSs is
potentially invalid in the sense of having a higher Type~I
error than nominal, that is, rejecting more often than desired under
Neyman's null.

However, the expected mean residual sum of squares is not necessarily
less than the expected mean treatment sum of squares
under Neyman's null. In fact, the inequality could go in either
direction. We describe in Section~\ref{sec:interactions_EMS} how the inequality depends on interactions
between row/column blocking factors and the
treatment.

%t2 #&#
%t2 ###
\begin{table}[b]
\caption{Table of potential outcomes for a LS with $\mathbb{E}(S_0^2) >
\mathbb{E}(S_1^2)$}\label{tab2}
\begin{tabular*}{\columnwidth}{@{\extracolsep{\fill}}lccc@{}}
\hline
& \textbf{Column} $\bolds{1}$ & \textbf{Column} $\bolds{2}$ & \multicolumn{1}{c@{}}{\textbf{Column} $\bolds{3}$} \\
\hline
Row $1$ & $(3,10,15)$ & $(50,30,13)$ & $(20,20,40)$ \\
Row $2$ & $(10,13,50)$ & $(20,40,3)$ & $(30,15,20)$ \\
Row $3$ & $(13,3,20)$ & $(15,20,10)$ & $(40,50,30)$ \\
\hline
\end{tabular*}
\end{table}

Two examples of LSs with $T = 3$, $\sigma_{\varepsilon}^2 = 0$, and
$\bar{X}_{\cdot\cdot}(\mathrm{1}) = \bar{X}_{\cdot\cdot}(\mathrm{2})
= \bar{X}_{\cdot\cdot}(\mathrm{3})$ (i.e., Neyman's
null) demonstrate this fact. In Tables~\ref{tab2} and \ref{tab3}, each
unit's potential outcomes are represented by an
ordered triple, with the $t$th coordinate denoting the
potential outcome under treatment $t$. For Table~\ref{tab2}, $\mathbb{E}(S_0^2) = 252.07, \mathbb{E}(S_1^2) = 172.38$.
From our formulae,
\begin{eqnarray*}
&&\mathbb{E}\bigl(S_0^2\bigr) - \mathbb{E}
\bigl(S_1^2\bigr) \\
&&\quad=-\frac{1}{(T-1)^2}\sum
_{t=1}^T \sigma_{\eta}^2(t) \\
&&\qquad{}+
\frac{1}{(T-1)^3}\sum_{t \neq t'} r\bigl(t,t'
\bigr)\sqrt{\sigma_{\eta}^2(t) \sigma_{\eta}^2
\bigl(t'\bigr)}
\\
& &\qquad{}+ \frac{1}{T(T-1)^2}\sum_{i=1}^T \sum
_{j=1}^T \sum_{t=1}^T
\bigl[\bigl\{ R_i(t) - \bar{R}_i(\cdot) \bigr
\}^2 \\
&&\hspace*{111pt}\qquad{}+ \bigl\{C_j(t) - \bar{C}_j(\cdot)
\bigr\}^2\bigr].
\end{eqnarray*}

We verify by explicit randomization that the discrepancy $\mathbb
{E}(S_0^2) - \mathbb{E}(S_1^2) = 79.69$ equals this
expression, so that this is one LS for which the expected mean residual
sum of squares is greater than the expected mean
treatment sum of squares. The inequality \vspace*{1pt} is in the other direction for
Table~\ref{tab3}, with
$\mathbb{E}(S_0^2) = 4.96, \mathbb{E}(S_1^2) = 6.77$.

%t3 #&#
%t3 ###
\begin{table}
\caption{Table of potential outcomes for a LS with $\mathbb{E}(S_0^2) <
\mathbb{E}(S_1^2)$}\label{tab3}
\begin{tabular*}{\columnwidth}{@{\extracolsep{\fill}}lccc@{}}
\hline
& \textbf{Column} $\bolds{1}$ & \textbf{Column} $\bolds{2}$ &
\multicolumn{1}{c@{}}{\textbf{Column} $\bolds{3}$} \\
\hline
Row $1$ & $(7,4,8)$ & $(5,9,4)$ & $(6,6,5)$ \\
Row $2$ & $(8,5,6)$ & $(3,3,3)$ & $(2,2,7)$ \\
Row $3$ & $(1,8,2)$ & $(4,7,9)$ & $(9,1,1)$ \\
\hline
\end{tabular*}
\end{table}

%s2.4 #&#
%s2.4 ###
\subsection{Latin Square Designs: After the Controversy}
\label{sec:LSD_history}

As with RCBs, no additivity assumption is made on the potential
outcomes for LSs. Nonadditivity for LSs has been further
studied in the literature (\citep{Gourlay1955b}; \citep{Tukey1955};
\citep{Rojas}). Kempthorne recognized the issue of interactions between
row/column blocking factors and the treatment factor in a LS (discussed
in the next section):

\begin{quote}
It is clear that, if there are row-treatment or column-treatment
interactions, these will enter into the error mean square
but not into the treatment mean square. The situation is entirely
analogous to that of randomized blocks in that
block-treatment interactions enter the error mean square but not the
treatment mean square.
(\citep{Kempthorne1952}, page 195)
\end{quote}

\noindent \citeauthor{Kempthorne1952} [(\citeyear{Kempthorne1952}), page 204] continued by noting a defect of large
LSs, namely, that there are more opportunities for
row/column interactions with treatments.

A substantial investigation in the spirit of Neyman was perfomed by
\citet{WilkKempthorne}, and is briefly summarized by
Hinkelmann and Kempthorne [(\citeyear{Kempthorne}), page 387].
Wilk and Kempt\-horne [(\citeyear{WilkKempthorne}), page 224] adopt the same specification of
potential outcomes as
\citet{Neyman1935}, allowing technical errors to differ based on
treatment level $k$:
\[
y_{ijk} = Y_{ijk} + \varepsilon_{ijk}.
\]
One difference that makes the conceptual yield framework of Wilk and
Kempthorne more general is that they consider randomly
sampling rows, columns and treatments from some larger population. In
any case,
\citeauthor{WilkKempthorne} [(\citeyear{WilkKempthorne}), page 227] reach the
reverse conclusion as Neyman, stating that, usually, the expected mean
residual sum of squares is larger than the expected
mean treatment sum of squares.
\citeauthor{WilkKempthorne} [(\citeyear{WilkKempthorne}), page 227]
explain this difference and the fact that Neyman did not
recognize interactions between row/column blocking factors and the
treatments, by noting that
\citeauthor{Neyman1935} [(\citeyear{Neyman1935}), page~145]
made additional homogeneity assumptions. However, Neyman's assumptions
were invoked solely to facilitate calculations and
had no physical justifications.

Our results are in agreement with a summary of their work in Table~3
from \citeauthor{WilkKempthorne} [(\citeyear{WilkKempthorne}), page 226]. Thus, it appears
that Wilk and Kempthorne do not seriously consider the possibility that
the inequality could go in the direction Neyman
claimed. In fact, \citeauthor{Kempthorne} [(\citeyear{Kempthorne}), page~387] when summarizing this
paper, explicitly state that the expected mean residual
sum of squares is larger than the expected mean treatment sum of
squares under Neyman's null. A possible explanation can be
found in the sixth remark on page $227$, where Wilk and Kempthorne
discuss how the standard approach to designing LSs may
likely result in interactions of row/column blocking factors with
treatments. As explained in our next section, the
magnitudes of these interactions ultimately drive the direction of the
inequality.

\citet{Cox1958} built on the work of Wilk and Kempt\-horne, and provided
a rather unique viewpoint on this entire
problem.\vadjust{\goodbreak} After first summarizing Wilk and Kempthorne's results by
stating that it is usually the case that the expected mean
residual sum of squares is larger than the expected mean treatment sum
of squares, Cox then considered the practical
importance of this difference of expectations, which he correctly
recognized as being related to interactions between the
treatment and blocking factors.
\citeauthor{Cox1958} [(\citeyear{Cox1958}), page 73] raised the thought-provoking question of
whether, for a LS, the practical scientific interest of the null
\[
H_0\dvtx\mathbb{E}\bigl(S_0^2\bigr) =
\mathbb{E}\bigl(S_1^2\bigr)
\]
is comparable to, or greater than, Neyman's null, especially when the
difference between these expected mean sums of squares
is considered important. He concluded that testing Neyman's null when
there is no unit-treatment additivity does not seem to
be helpful:

\begin{quote}
\ldots if substantial variations in treatment effect from unit to unit do
occur, one's understanding of the experimental
situation will be very incomplete until the basis of this variation is
discovered and any extension of the conclusions to a
general set of experimental units will be hazardous. The mean treatment
effect, averaged over all units in the experiment,
or over the finite population of units from which they are randomly
drawn, may in such cases not be too helpful.
Particularly if appreciable systematic treatment-unit interactions are
suspected, the experiment should be set out so these
may be detected and explained. (\citep{Cox1958}, page 73)
\end{quote}

\citeauthor{Cox2012} [(\citeyear{Cox2012}), page 3] later argued that when this more realistic
null is formulated, the biases described earlier disappear,
and so do issues surrounding the LS. A related point for the LS design
noted by Cox is the marginalization principle, in
which models having nonzero interactions and zero main effects are not
considered sensible [similar to the effect heredity
principle (\cite{wuhamada}, page 173)]. \citet{Box1984}, when commenting
on \citet{Cox1984}, provided an opposing view that
makes such a principle context-dependent.

%s2.5 #&#
%s2.5 ###
\subsection{Block-Treatment Interactions and Expected Sums of Squares}
\label{sec:interactions_EMS}

Neyman excluded the following (respective) terms in $\mathbb{E}(S_0^2)$
for RCBs and LSs:
\begin{eqnarray*}
&&\frac{1}{(N-1)(T-1)} \sum_{i=1}^N \sum
_{t=1}^T \bigl\{ B_i(t) - \bar
{B}_i(\cdot) \bigr\}^2,
\\
&&\frac{1}{(T-1)^2}\sum_{i=1}^T\sum
_{t=1}^T \bigl\{ R_i(t) -
\bar{R}_i(\cdot) \bigr\}^2 \\
&&\quad{}+ \frac{1}{(T-1)^2}\sum
_{j=1}^T\sum_{t=1}^T
\bigl\{ C_j(t) - \bar{C}_j(\cdot) \bigr
\}^2.
\end{eqnarray*}
In each, we are adding squared differences between the fertility
correction for a specific combination of block and
treatment levels, and the average (over treatments) fertility
correction for the same block level. For the LS, this is
decomposed as a sum over the row and a sum over the column blocking factors.

Formally, these terms gauge whether, for each level of a blocking
factor, the fertility corrections are constant over the
treatments, and represent interactions between blocking factors and
treatments. For RCBs, we have
\[
B_i(t) - \bar{B}_i(\cdot) = \bigl\{
\bar{X}_{i \cdot}(t) - \bar{X}_{i \cdot}(\cdot) \bigr\} - \bigl\{ \bar
{X}_{\cdot\cdot}(t) - \bar{X}_{\cdot\cdot}(\cdot)\bigr\},
\]
which is the interaction between the $i$th block and the $t${th}
treatment in terms of potential
outcomes. Similarly, we have for LSs that
\begin{eqnarray*}
R_i(t) - \bar{R}_i(\cdot) &=& \bigl\{
\bar{X}_{i \cdot}(t) - \bar{X}_{i \cdot}(\cdot) \bigr\} - \bigl\{ \bar
{X}_{\cdot\cdot}(t) - \bar{X}_{\cdot\cdot}(\cdot)\bigr\},
\\
C_j(t) - \bar{C}_j(\cdot) &=& \bigl\{
\bar{X}_{\cdot j}(t) - \bar{X}_{\cdot j}(\cdot) \bigr\} - \bigl\{ \bar
{X}_{\cdot\cdot}(t) - \bar{X}_{\cdot\cdot}(\cdot)\bigr\},
\end{eqnarray*}
which are the interactions between the $i$th row and $t${th}
treatment, and the $j$th column and
the $t${th} treatment, respectively, in terms of potential outcomes.

Intuitively, these interactions, which are functions of potential
outcomes, should reside within the expectation of the mean
residual sum of squares. Without invoking additivity on the potential
outcomes, these interactions are not necessarily zero
and, because we lack replications within blocks for either RCB or LS
designs, we cannot form an interaction sum of squares
from the observed data, so that the potential outcome interactions will
instead be included in the expectation of the mean
residual sum of squares (\citep{Fisher}, Chapters IV, V). In contrast,
for randomized block designs that include replications
within each block, this interaction term is no longer present in the
expected mean residual sum of squares.

To better understand the expected mean sums of squares for LSs,
consider their difference under Neyman's simplifying
assumption that $\sigma_{\eta}^2(t)$ and $r(t,t')$ are constant, so that
$\sigma_{\eta}^2(t) = \sigma_{\eta}^2$ and $r(t,t') = r$ for all
treatments $t, t'$. Then the difference between
$\mathbb{E}(S_0^2)$ and $\mathbb{E}(S_1^2)$ under Neyman's null is
\begin{eqnarray*}
&&\sum_{i=1}^T \sum
_{t=1}^T\bigl\{R_i(t) -
\bar{R}_i(\cdot)\bigr\}^2 \\
&&\quad{}+ \sum
_{j=1}^T \sum_{t=1}^T
\bigl\{C_j(t) - \bar{C}_j(\cdot)\bigr\}^2 -
T\sigma_{\eta}^2(1 - r),
\end{eqnarray*}
and this expression, in some sense, measures the difference between
row/column interactions with treatment and the variance
of the potential outcome residual terms (scaled by the number of
treatments, $T$, times one minus the correlation between
potential outcome residual terms for different pairs of treatments).
Note that
$0 \leq1 - r \leq2$, so $0 \leq T\sigma_{\eta}^2(1-r) \leq2T\sigma
_{\eta}^2$.

To interpret the difference in expectations for the general case, first
note that
\begin{eqnarray*}
\sum_{i=1}^T \sum
_{j=1}^T\bar{\eta}_{ij}(
\cdot)^2 &\geq& 0\quad \Rightarrow\\
 \sum_{t=1}^T
\sigma_{\eta}^2(t) &\geq&-\sum_{t \neq t'}
r\bigl(t,t'\bigr)\sqrt {\sigma_{\eta}^2(t)
\sigma_{\eta}^2\bigl(t'\bigr)}.
\end{eqnarray*}
As such, $\mathbb{E}(S_0^2) - \mathbb{E}(S_1^2)$ under Neyman's null is
bounded from below by
\begin{eqnarray*}
&&\frac{1}{(T-1)^2}\sum_{i=1}^T \sum
_{t=1}^T\bigl\{R_i(t) -
\bar{R}_{i}(\cdot )\bigr\}^2\\
&&\quad{} + \frac{1}{(T-1)^2}\sum
_{j=1}^T \sum_{t=1}^T
\bigl\{C_j(t) - \bar{C}_{j}(\cdot )\bigr\}^2 \\
&&\quad{}-
\frac{T}{(T-1)^3}\sum_{t=1}^T
\sigma_{\eta}^2(t),
\end{eqnarray*}
so that, if
\begin{eqnarray*}
&&\sum_{i=1}^T \sum
_{t=1}^T\bigl\{R_i(t) -
\bar{R}_{i}(\cdot)\bigr\}^2 + \sum
_{j=1}^T \sum_{t=1}^T
\bigl\{C_j(t) - \bar{C}_{j}(\cdot)\bigr\}^2\\
&&\quad{} -
\frac{T}{T-1}\sum_{t=1}^T
\sigma_{\eta}^2(t) \geq0,
\end{eqnarray*}
then $\mathbb{E}(S_0^2) \geq\mathbb{E}(S_1^2)$.
Even in the most general case for LSs, $\mathbb{E}(S_0^2) - \mathbb
{E}(S_1^2)$ can still be interpreted as a comparison
between row/column interactions with treatment and the (scaled) sum of
variances of residual potential outcomes
$\eta_{ij}(t)$.

In the context of an agricultural experiment, we obtain a more
meaningful interpretation for this difference. Latin squares
are implemented to block on fertility gradients in two direction
(\citep{Neyman1935}; \citep{Fisher}, Chapter V;
Hinkelmann and Kempthorne, \citeyear{Kempthorne}, Chapter~10). If the variability of specific soil
fertility corrections across rows and columns (i.e.,
interactions between rows/columns and treatments) are substantially
larger than the residual variability of the potential\vadjust{\goodbreak}
outcomes [i.e., the variability of the $\eta_{ij}(t)$], then $\mathbb
{E}(S_0^2) - \mathbb{E}(S_1^2)$ is larger than zero. An
example was given in Table~\ref{tab2}, where
\begin{eqnarray*}
&&\sum_{i=1}^T \sum
_{t=1}^T\bigl\{R_i(t) -
\bar{R}_i(\cdot)\bigr\}^2 + \sum
_{j=1}^T \sum_{t=1}^T
\bigl\{C_j(t) - \bar{C}_j(\cdot)\bigr\}^2\\
&&\quad=
569.93,
\\
&&-\sum_{t=1}^T \sigma_{\eta}^2(t)
= -313.56,
\\
&&\frac{1}{T-1}\sum_{t \neq t'} r\bigl(t,t'
\bigr)\sqrt{\sigma_{\eta}^2(t) \sigma
_{\eta}^2\bigl(t'\bigr)} = 62.41.
\end{eqnarray*}
The interaction is nearly twice the variability of the residual
potential outcomes, and so the difference
$\mathbb{E}(S_0^2) - \mathbb{E}(S_1^2)$ is greater than zero. For Table~\ref{tab3},
\begin{eqnarray*}
&&\sum_{i=1}^T \sum
_{t=1}^T\bigl\{R_i(t) -
\bar{R}_i(\cdot)\bigr\}^2 + \sum
_{j=1}^T \sum_{t=1}^T
\bigl\{C_j(t) - \bar{C}_j(\cdot)\bigr\}^2 \\
&&\quad=
9.48,
\\
&&-\sum_{t=1}^T \sigma_{\eta}^2(t)
= -14.59,
\\
&&\frac{1}{T-1}\sum_{t \neq t'} r\bigl(t,t'
\bigr)\sqrt{\sigma_{\eta}^2(t) \sigma
_{\eta}^2\bigl(t'\bigr)} = -2.11,
\end{eqnarray*}
and the variance of the residuals completely dominates the interaction.

Hence,
$\mathbb{E}(S_0^2) > \mathbb{E}(S_1^2)$ in the presence of a strong
fertility gradient, with the
interaction between row/column blocking factors and treatment greater
than the variance of the residual potential
outcomes or, alternatively, when the unit-treatment interactions are
negligible. Similarly,
$\mathbb{E}(S_0^2) < \mathbb{E}(S_1^2)$ in cases where no strong
interaction exists between row/column blocking factors and
the treatment when compared to the variability of the residual
potential outcomes or, alternatively, when the unit-treatment
interactions are substantial. It is important to recognize that such
important interactions can never be assessed without
replication, which is not available in the original LS design.

%s3 #&#
%s3 ###
\section{Controversial Connections}
\label{sec:3}

%s3.1 #&#
%s3.1 ###
\subsection{Connecting Expected Mean Sums of Squares with Type I Error}
\label{sec:cacophony}

\citeauthor{Neyman1935} (\citeyear{Neyman1935}) calculated expectations of mean sums of squares to
argue that the standard ANOVA F-test for RCB designs
is valid and the test for LS designs is invalid when testing
Neyman's\vadjust{\goodbreak}
null: a test was said to be ``unbiased'' if
$\mathbb{E}(S_0^2) = \mathbb{E}(S_1^2)$ under Neyman's null
(\citep{Neyman1935}, page 144). The reasoning behind this definition
is not discussed at all and, given our current understanding of
hypothesis testing, seems somewhat crude. After
all, to determine whether a particular testing procedure is ``biased,''
one typically calculates the probability of
rejecting a true null hypothesis, which generally depends on the test
statistic's distribution, not just its expectation.

To better understand the logic potentially driving Neyman's reasoning,
it is useful to review the testing of Fisher's sharp
null. A randomization test that uses any {a priori} defined test
statistic automatically yields the correct Type I
error under Fisher's sharp null and regularity conditions on the
potential outcomes and number of randomizations.
Furthermore, when using the statistic $F = S_1^2/S_0^2$, this
randomization distribution is well approximated by the
F-distribution, for both RCB and LS designs. \citet{Welch} calculated
the first two moments of
%
%e3.1 #&#
%e3.1 ###
\begin{equation}
\frac{df_1 S_1^2}{df_1 S_1^2 + df_0 S_0^2} = \frac{df_1 F}{df_1 F + df_0}, \label{eq:trick}
\end{equation}
where $df_1$ denotes the degrees of freedom for treatment sum of
squares, and $df_0$ the degrees of freedom for residual sum
of squares. \citet{Pitman} calculated the first four moments of this
statistic. For both RCB and LS designs,
$df_1 S_1^2 + df_0 S_0^2$ remains constant over the randomizations
under Fisher's sharp null, making calculation
of the moments of (\ref{eq:trick}) much easier than of $F$ itself.
Furthermore, under regularity conditions on the potential
outcomes, it was shown that these moments are approximately equal to
the corresponding moments of a Beta distribution. In
this respect, the standard ANOVA F-test that uses rejection cutoffs
based on the F-distribution has approximately the
correct Type I error, and the F-distribution can be viewed as a simple
approximation to the randomization distribution of
the F-test statistic when testing Fisher's sharp null
(\citep{Kempthorne1952}, pages 172, 193). Indeed, as stated by
\citeauthor{Wilk1955} [(\citeyear{Wilk1955}), page 77] the amount of computation to perform a
randomization test could be prohibitive, and statisticians
had little recourse except to use such approximations. Kempthorne made
a similar remark:

\begin{quote}
It should be realized that the analysis of variance test with the F
distribution has a fair basis apart from normal law
theory and is probably in most cases a good approximation to the
randomization analysis of variance test, which is a
nonparametric test. (\citep{Kempthorne1955}, page 966)
\end{quote}

Kempthorne earlier stated that for LSs:

\begin{quote}
The randomization test for the Latin Square or for any randomized
design is entirely valid in the sense of controlling Type
I errors, but the approximation to this test by the F-distribution when
there is nonadditivity is apparently completely
unknown. (\citep{Kempthorne1955}, page 965)
\end{quote}

\noindent As Neyman did not invoke additivity or any other regularity conditions
on the potential outcomes, the reasoning outlined in
the previous paragraph that establishes the F-distribution as an
approximation to the true distribution of the F-test
statistic is no longer valid when testing Neyman's null: for example,
$df_1 S_1^2 + df_0 S_0^2$ is generally no longer constant
over the randomizations, and calculating moments of equation~(\ref
{eq:trick}) generally becomes very difficult.
\citeauthor{Wilk1955} [(\citeyear{Wilk1955}), page~79] realized this, remarking that the standard
ANOVA F-test for testing Neyman's null in RCBs depends on
the assumption that block-treatment interactions are zero.
\citeauthor{WilkKempthorne} [(\citeyear{WilkKempthorne}), page~228] also stated that the effect of
nonadditivity on the Type I error of the standard ANOVA F-test for a LS
is unknown.

Bearing these facts in mind, a comparison of expected mean residual and
treatment sums of squares could be viewed as a crude
way of assessing whether the Type I error is correct when testing
Neyman's null using the standard ANOVA F-test.
\citet{Neyman1935} himself may have realized this:

\begin{quote}
\ldots in the case of the Randomized Blocks the z test may be considered
as unbiased in the sense that the
expectations of $S_0^2$ and $S_1^2$ have a common value \ldots On the
other hand,\vadjust{\goodbreak} by the arrangement in Latin Square
the expectation of $S_1^2$ is equal to $\frac{1}{2}n'\sigma_d^2$, while
that of $S_0^2$ is generally smaller. This
suggests, although it does not prove, that by the Latin Square
arrangement the z test may have the tendency to detect
differentiation when it does not exist. (\citep{Neyman1935}, page 144)
\end{quote}

\noindent After calculating expected mean sums of squares for RCBs, Neyman states that

\begin{quote}
If there is no differentiation among the $X_{\cdot\cdot}(k)$, then
$\mathbb{E}(S_1^2) = \mathbb{E}(S_0^2)$,\vadjust{\goodbreak}
and we see that the test of significance usually applied is unbiased in
the sense that if there is no
differentiation, then the values of $S_1^2$ and $S_0^2$ must be
approximately equal. This, of course, does not prove
the validity of Fisher's z test. (\citep{Neyman1935}, page 150)
\end{quote}

\noindent Furthermore, Neyman states that for LSs:

\begin{quote}
We conclude, therefore, that at present there is no theoretical
justification for the belief that the z test
is valid in the case of the arrangement by the Latin Square: not only
is there the difficulty connected with the
nonnormality of the distribution of the $\eta$'s, but also the
functions which are usually considered as unbiased
estimates of the same variance have generally different expectations.
This may (though not necessarily so) cause a
tendency to state significant differentiation when this, in fact, does
not exist. \ldots These, of course, are purely
theoretical conclusions, and I am personally inclined to think that
from the practical point of view the existing bias
will prove to be negligible. (\citep{Neyman1935}, page 154)
\end{quote}

This same consideration of expected mean sums of squares for hypothesis
testing continues in the present literature
on experimental design:

\begin{quote}
It is the form of the expected mean squares, $\mathbb{E}[\mathrm
{MS}(i)]$, which determines, for example, how
tests of hypotheses are performed and how error variances are
estimated.
(\citep{Kempthorne}, page~37)
\end{quote}

\noindent Also:

\begin{quote}
In this case, MS(E) is on average larger than MS(T) under the
hypothesis of no treatment effects and hence
the usual F-test will lead to fewer significant results. In this case
the LSD is not an unbiased design.
(\citep{Kempthorne}, page~387)
\end{quote}

It is interesting to note that the specific justification for this
last statement was never made, nor
was any attempt made to calculate explicitly the Type I error. Even
more interesting is how these statements contradict
Kempthorne's earlier\vadjust{\goodbreak} position on the connection between expected mean
sums of squares and hypothesis testing
(e.g., as given by \citeauthor{Kempthorne1952} [(\citeyear{Kempthorne1952}),
page 149]), for example:

\begin{quote}
To establish the property of unbiasedness for this design it is \ldots
necessary to show that the expectation over
randomizations of the error mean square resulting from this model is
equal to the mean square among all observations in the
absence of treatment effects. \ldots it should perhaps be noted that this
property has no intrinsic relation to the
concept of unbiasedness of a test. (\citep{Kempthorne1955}, page 956)
\end{quote}

\noindent \citet{WilkKempthorne} hold this same position, stating that:

\begin{quote}
We accept the view that tests of significance are evaluatory procedures
leading to assessments of strength of evidence
against particular hypotheses, while tests of hypotheses are decision
devices. We are here concerned with the former, and in
this connection it should be noted that (a) the expectations of mean
squares are in some degree irrelevant to the exact
(permutation) test of significance of the null hypothesis that the
treatments are identical.
(\citep{WilkKempthorne}, page 228)
\end{quote}

%s3.2 #&#
%s3.2 ###
\subsection{Concrete Calculations}
\label{sec:concrete}

From Section~\ref{sec:RCBD_theory}, the F-test for RCBs is generally
biased in one direction under Neyman's conception of an
unbiased test, potentially leading to fewer rejections under Neyman's
null. Furthermore, because we do not make any
assumptions about the difference between the interactions of
rows/columns with treatment and the residual variances in
Section~\ref{sec:LSD_theory}, we actually cannot claim that the F-test
for LSs is biased in any one direction. A more
rigorous justification for the ``unbiasedness'' of the F-test for
either design would compare the actual distribution of the
F-test statistic to the associated F-distribution. By determining
whether the distribution of $F = S_1^2/S_0^2$ is
adequately approximated by the F-distribution under Neyman's null, one
would be able to conclude whether the Type I error is
approximately as advertised.

%t4 #&#
%t4 ###
\begin{table}[b]
\caption{Table of potential outcomes for a $4 \times4$ LS, with
$\mathbb{E}(S_0^2) = \mathbb{E}(S_1^2)$}\label{tab4}
\begin{tabular*}{\columnwidth}{@{\extracolsep{\fill}}lcccc@{}}
\hline
& \textbf{Column} $\bolds{1}$ & \textbf{Column} $\bolds{2}$ &
\textbf{Column} $\bolds{3}$ & \multicolumn{1}{c@{}}{\textbf{Column} $\bolds{4}$} \\
\hline
Row $1$ & $(1,1,1,1)$ & $(0,0,0,0)$ & $(0,0,0,0)$ & $(0,0,0,0)$ \\
Row $2$ & $(0,0,0,0)$ & $(1,1,1,1)$ & $(0,0,0,0)$ & $(0,0,0,0)$ \\
Row $3$ & $(0,0,0,0)$ & $(0,0,0,0)$ & $(0,0,0,0)$ & $(0,0,0,0)$ \\
Row $4$ & $(0,0,0,0)$ & $(0,0,0,0)$ & $(0,0,0,0)$ & $(0,0,0,0)$ \\
\hline
\end{tabular*}
\end{table}

%f1 #&#
%f1 ###
\begin{figure*}[b]

\includegraphics{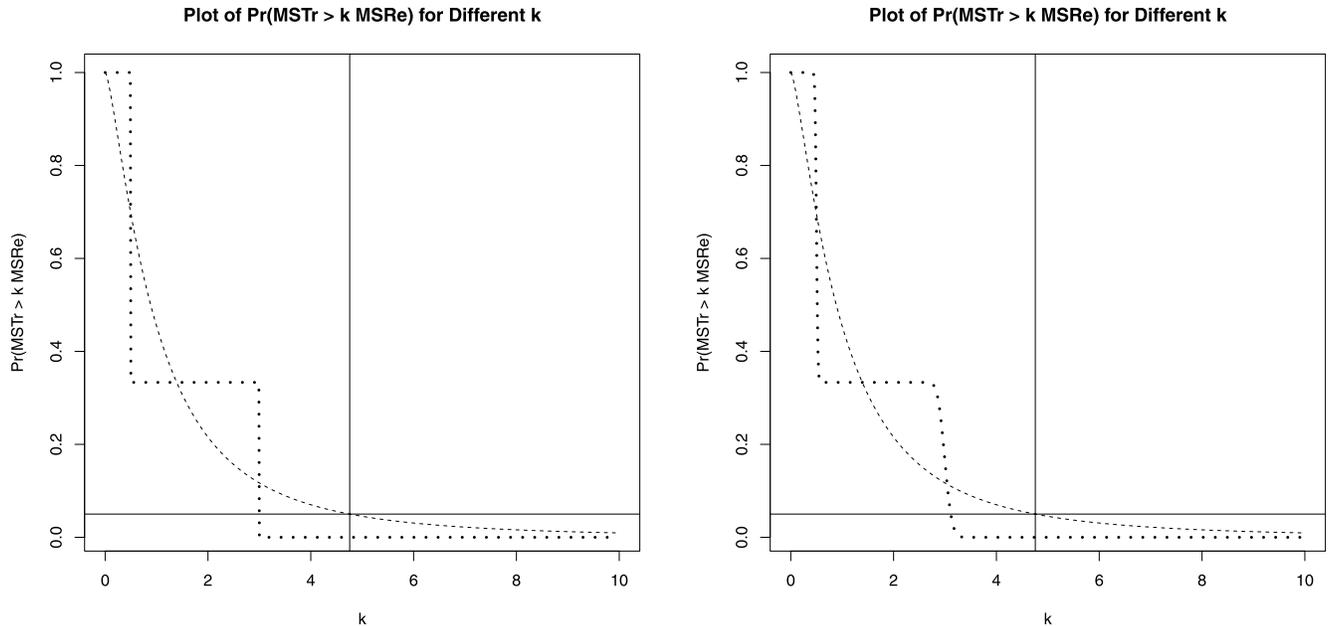}

\caption{Comparison of the distributions of $S_1^2/S_0^2$ and $F_{3,6}$ for
Table \protect\ref{tab4};
the distribution of $S_1^2/S_0^2$ is represented by dots and that
of $F_{3,6}$ by dashes.
The figure on the left is for the case with no technical errors, and
the figure on the right
is for technical errors with $\sigma_{\varepsilon} = 0.01$.}
\label{fig1}
\end{figure*}

We performed this comparison for various RCBs and LSs, and observed
that Neyman's definition of unbiased tests fails. In
particular, we can generate infinitely\vadjust{\goodbreak} many RCBs and LSs such that
$(1)$ Neyman's null holds, $(2)$ there is no interaction
between blocking factor(s) and treatment, $(3)$ the expected mean
residual sum of squares equals the expected mean treatment
sum of squares, and yet there is zero probability of rejecting Neyman's
null when the rejection rule is based on a
comparison of the observed value of $S_1^2/S_0^2$ with $\alpha= 0.05$
cutoffs used in the standard ANOVA F-test.

For simplicity, consider the case with no technical errors. One simple
example of a $4 \times4$ LS,
with $\sigma_{\eta}^2(t), r(t,t')$ constant, $\mathbb{E}(S_0^2) =
\mathbb{E}(S_1^2)$, and no interactions between
row/column blocking factors and the treatment, is presented in Table~\ref{tab4}. Now $F_{3,6,0.95} = 4.76$ and, as we have
all potential outcomes, we can calculate the probability that $S_1^2 >
k S_0^2$ for any positive number $k$ over the
distribution of $S_1^2$ and $S_0^2$. These probabilities are given in
the left of Figure~\ref{fig1}, which also displays
probabilities that $F_{3,6} > k$; probabilities from the randomization
distribution of $S_1^2/S_0^2$ are plotted as dots,
and probabilities for the $F_{3,6}$ distribution as dashes.
A~horizontal line at $0.05$ and a vertical line at $4.76$ were
drawn to illustrate conclusions obtained at the $0.05$ significance
level. The probability of rejecting Neyman's null when
using the standard ANOVA F-test is zero.

The crucial factor here is the structure of the potential outcomes.
Fisher's sharp null holds, so the total sum of squares,
and the sum of squares for row and column blocking factors, remain
constant over the randomization. Furthermore, the
treatment sum of squares takes only two values, corresponding to
whether cells $(1,1)$ and $(2,2)$ receive the same
treatment or not, and similarly the residual sum of squares takes only
two values. Hence, the F-test statistic takes only
two possible values, so that cutoffs given by consideration of the
F-distribution will not yield approximately correct Type
I errors for testing Neyman's null.

Inclusion of technical errors does not change our general conclusion.
Suppose technical errors are Normally distributed with
$\sigma_{\varepsilon} = 0.01$. The corresponding figure for the LS in
Table~\ref{tab4} is displayed in the right of Figure~\ref{fig1}. We generated this figure by simulation: we first drew
$\varepsilon_{ij}(t)$, then performed the randomizations to
generate the distribution of $S_1^2$ and $S_0^2$ for that specific draw
of technical errors, and finally repeated this
process $2000$ times to estimate the probabilities.

%s4 #&#
%s4 ###
\section{Controversial Consequences and Conclusions}
\label{sec:4}

%s4.1 #&#
%s4.1 ###
\subsection{Consequences}
\label{sec:consequences}

The most immediate consequence of this entire controversy was the
resulting hostile relationship between Neyman and Fisher
for essentially the remainder of their careers, with each seeking to
undermine the other. For example, Neyman was slightly
critical in a discussion of a paper presented by \citet{Yates1935} on
factorial designs.
\citeauthor{Box} [(\citeyear{Box}), page 265] claimed that
Neyman wanted to demonstrate his superiority by finding flaws in
Fisher's work at this meeting.
\citeauthor{Reid} [(\citeyear{Reid}), page 126]
described an interesting encounter between Neyman and Fisher, taking
place in Neyman's room at University College London
one week after this discussion. Fisher demanded that Neyman only use
Fisher's books when lecturing on statistics at the
university. When Neyman refused to do so, Fisher openly declared that
he would oppose Neyman in all his capacities, and
banged the door when he left the room.

These skirmishes continued for some time
(\citep{Reid}, pages 143, 169, 183--184, 223--226, 256--257). Neyman
appears to have
attempted some type of reconciliation, inviting Fisher to lecture at
Berkeley
(\citep{Reid}, page 222), and generally became
more conciliatory toward Fisher and his contributions to statistics
(\citep{Neyman1976}; \citep{Reid}, page 45). In any
case, these passages suggest an indirect consequence of this
controversy: Neyman's decision to depart for America, where he
created a world-class center for statistics at the University of
California Berkeley
(\citep{Reid}, page 239), established a
prominent series of symposia (\citep{Reid}, pages 197--198), and helped
to nurture, through his leadership, the American
Statistical Association and Institute of Mathematical Statistic
(\citep{Reid}, page 218).

\citet{FienbergTanur1996} suggest that this break in the professional
relationship between Neyman and Fisher may have led to a sharper
division between the fields of sample surveys and experimental design:

\begin{quote}
Because of the bitterness that grew out of this dispute \ldots Fisher
and\vadjust{\goodbreak}
Neyman were never able to bring their ideas together
and benefit from the fruitful interaction that would likely have
occurred had they done so. And in the aftermath, Neyman
staked out intellectual responsibility for sampling while Fisher did
the same for experimentation. It was in part because
of this rift between Fisher and Neyman that the fields of sample
surveys and experimentation drifted apart.
(\citep{FienbergTanur1996}, page 238)
\end{quote}

\citet{Cox2012} makes the interesting remark that more effort was
devoted to issues in randomization following this
controversy:

\begin{quote}
The general issues of the role of randomization were further discussed
in the next few years, mostly in
\textit{Biometrika}, with contributions from Student, Yates, Neyman and
Pearson, and Jeffreys. With the exception of
Student's contribution, which emphasized the role of randomization in
escaping biases arising from personal judgement, the
discussion focused largely on error estimation. (\citep{Cox2012}, page 3)
\end{quote}

Another consequence was undue emphasis on linear models for analysis of
experimental data. As stated by
\citeauthor{Gourlay1955a} [(\citeyear{Gourlay1955a}), page 228] Neyman's
work in $1935$ led to
increased attention on models (for observed data) that formed
the basis of statistical analyses such as ANOVA. \citet{Eisenhart1947},
for example, explicitly laid out
the four standard assumptions used to justify ANOVA, and noted the
importance of additivity. Immediately following this
article, \citet{Cochran1947} explored the consequences for an analysis
when additivity (and the other assumptions) were not
satisfied, and \citet{Bartlett1947} discussed various transformations
of the data that make additivity more plausible for
ANOVA.

Accordingly, past and present books on experimental design tend to
invoke additive models when testing Neyman's null using
the standard ANOVA F-test, an assumption that automatically yields a
test of Fisher's sharp null
(\citep{Kempthorne1952}, Chapters~8, 9, 10;
\citep{Kempthorne}, Chapters~9, 10). When additivity is believed not to
hold, one
is generally advised to search for a transformation that yields an
additive structure on the potential outcomes. For
example, \citeauthor{WilkKempthorne} [(\citeyear{WilkKempthorne}), page 229] make the strong
recommendation to\vadjust{\goodbreak} transform to a scale where additivity more nearly
obtains for purposes of estimation. This also reflects the motivation
behind the famous \citet{BoxCox} family of
transformations.

Of course, greater emphasis on linear models with Normal errors for
observed potential outcomes can generate doubts as to
whether randomization is necessary in experimental design. What is then
lost is the fact that explicit randomization,
as extolled by Fisher, provides the scientist with internally
consistent statistical inferences that require no standard
modeling assumptions, such as those required for linear regression. It
is ironic that many textbooks on experimental design
focus solely on Normal theory linear models, without realizing that
such models were originally motivated as approximations
for randomization inference.

Additivity has even been considered an essential assumption for
interpreting estimands. For example,
\citeauthor{Cox} [(\citeyear{Cox}), pages 16--17]
states that the average difference in observed outcomes for two
treatments estimates the difference in average potential
outcomes for the two treatments in the finite population, but that this
estimand of interest is \textit{``\ldots rather an
artificial quantity''} if additivity does not hold on the potential
outcomes. Perhaps
\citeauthor{Kempthorne1952} [(\citeyear{Kempthorne1952}), page 136] can
best justify this statement with the specific example where, for each
experimental unit, the square root of the potential
outcome under treatment is $5$ more than the square root of the
potential outcome under control. If one experimenter has
three experimental units with control potential outcomes equal to $25,
64$ and $100$, then the effect of the treatment on
the raw measurement scale would range from $75$ to $125$. However,
another experimenter working with units having control
potential outcomes ranging from $9$ to $16$ would have treatment
effects ranging from $55$ to $65$ on the raw scale. As
Kempthorne states:

\begin{quote}
Under these circumstances both experimenters will agree only if they
state their results in terms of effects on the square
root of the observation. It is desirable then to express effects on a
scale of measurement such that they are exactly
additive. (\citep{Kempthorne1952}, page 136)
\end{quote}

Thus, Kempthorne's justification for additivity is that it enables
externally consistent conclusions to be drawn from a
particular analysis, that is, two experimenters working with different
samples from the same population will reach the same\vadjust{\goodbreak}
conclusion on the treatment effect. One could also interpret this as
suggesting that experimenters should model the
potential outcomes, with additive treatment effects being one simple
model for an analysis.

Kempthorne continues to state that:

\begin{quote}
Such a procedure has its defects, for experimenters prefer to state
effects on a scale of measurement that is used as a
matter of custom or for convenience reasons. It is probably difficult,
for instance, to communicate to a farmer the meaning
of the statement that a certain dose of an insecticide reduces the
square root of the number of corn borers. A statement on
the effect of number of corn borers can be made but is more complex.
These difficulties are not, however, in the realm of
the experimenter. He should examine his data on a scale of measurement
which is such that treatment effects are additive.
The real difficulty, in general, is to determine the scale of
measurement that has the desired property.
(\citep{Kempthorne1952}, page 136)
\end{quote}

We again read in this quote the perceived importance of additivity that
helped motivate the \citet{BoxCox} family of
transformations. We do not believe it is necessary to study treatment
effects on an additive scale: it is arguably more
important to have an internally consistent definition and statistical
procedure for studying treatment effects before
deciding on externally consistent considerations. In our opinion, an
ultimate consequence of this controversy is that, by
focusing almost solely on linear models, advances in experimental
design have been seriously inhibited from their original,
useful and liberating formulation involving potential outcomes.

%s4.2 #&#
%s4.2 ###
\subsection{Conclusions}
\label{sec:conclusion}

The Neyman--Fisher controversy arose in part because Neyman sought to
determine whether Fisher's ANOVA F-test for RCBs and
LSs would still be valid when testing Neyman's more general null
hypothesis. Unfortunately, Neyman's calculations were
incorrect. In fact, under Neyman's conception of unbiased tests, the
F-test for RCB designs potentially rejects at most at
the nominal level, yet we could never know for any particular situation
whether the F-test for LS designs would reject more
often than nominal or not. Furthermore, Neyman's definition of unbiased
tests is too crude, because expected\vadjust{\goodbreak} mean sums of
squares do not determine the Type I error of the F-test when testing
Neyman's null. Two of the greatest statisticians argued
over incorrect calculations and inexact measures of unbiasedness for
hypothesis tests, adding an ironic aspect to this
controversy.

What is also ironic is that apparently no statistician deigned to check
Neyman's algebra or reasoning; the only discussant
who suggested there was a mistake in Neyman's algebra was Fisher, but
he did not explicitly state that Neyman was missing
interactions in both expected mean residual sums of squares.
\citeauthor{Sukhatme} [(\citeyear{Sukhatme}), pages 166, 167] recalculated the expected mean
sums of squares in the general case where $\sigma_{\eta}^2(t)$ and
$r(t,t')$ are not constant, and did not catch Neyman's
mistake. Sukhatme also performed sampling experiments for two examples
of LSs to support Neyman's claims. In both of
Sukhatme's examples, there is no interaction between row/column
blocking factors and treatment, so that
$\mathbb{E}(S_0^2) < \mathbb{E}(S_1^2)$.
\citeauthor{Neyman1935} [(\citeyear{Neyman1935}), page 175] then considered his algebra correct, because
\textit{`` \ldots none of my critics have attempted to challenge it.''}

Fisher never referenced \citet{Neyman1935} in his book on experimental
design and apparently ignored potential outcomes for
many years (\citep{Rubin2005};
\citep{Lehmann}, page 59). Fisher's avoidance of potential outcomes led
him to make certain
oversights in causal inference. In particular, as described by \citet
{Rubin2005}, Fisher never bridged his work on
experimental design and parametric modeling, and gave generally flawed
advice on the analysis of covariance to adjust for
posttreatment concomitants in randomized trials.

There is only one reference to \citet{Neyman1935} by
\citeauthor{Kempthorne} [(\citeyear{Kempthorne}), page 387] and it was referred to as
\textit{``\ldots an interesting somewhat different discussion \ldots''}. The
standard accounts of Fisher and Neyman's professional
careers (\citep{Box}; \citep{Reid}) do not mention any further work
being done on questions raised by \citet{Neyman1935}, although
Kempthorne is quoted as saying:

\begin{quote}
The allusion to agriculture is quite unnecessary and the discussion is
relevant to experimentation in any field of human
enquiry. The discussion section \ldots is interesting because of the
remarks of R. A. Fisher which are informative in some
respects but in other respects exhibit Fisher at his very worst \ldots.
The\vadjust{\goodbreak}
judgement of the future will be, I believe, that
Neyman's views were in the correct direction. (\citep{Reid}, page 123)
\end{quote}

Even the recent account by \citeauthor{Lehmann} [(\citeyear{Lehmann}), Chapters~4, 5]
does not
mention any statistician addressing Neyman's
claims or checking his algebra. In fact, Lehmann ends his discussion of
this controversy by recounting the destruction of
the physical models Neyman used to illustrate his thoughts on RCB and
LS designs during his $1935$ presentation, thought to
have been perpetrated by Fisher in a fit of anger
(\citep{Reid}, page 124; \citep{Lehmann}, Chapter~4).

We agree with Kempthorne's assessment that Neyman's views were in the
correct direction in the following sense: by
evaluating the frequency properties of statistics for both designs, one
can see that the F-test is no longer precise
without further assumptions on the potential outcomes. Such evaluations
serve the important task of investigating the
general properties of a design in a particular applied setting. The
F-distribution is a useful approximation to the
randomization distribution of the F-test statistic under Fisher's sharp
null hypothesis and regularity conditions on the
distribution of the potential outcomes or, alternatively, for testing
Neyman's null under additivity
(\citep{Welch}; \citep{Pitman}).

We also agree with \citet{Cox1958} that, if block-treatment
interactions are not negligible, then it is not particularly
useful to test Neyman's null. More generally, we believe that one must
think carefully about the type of null hypotheses
one will test, and should be guided by an appropriate model on the
potential outcomes. At one extreme, Fisher's sharp null
hypothesis requires no model on the potential outcomes to test a
reasonable, scientifically interesting null, with the
reference distribution based solely on the randomization actually
implemented during the experiment. To test Neyman's null,
one either needs strong regularity conditions on the potential outcomes
for standard procedures to work or one needs to
think carefully to build and evaluate a model for the potential
outcomes. In any case, one necessarily needs to make
assumptions to assess more complicated null hypotheses, and it is
important that assumptions on the potential outcomes are
driven by actual science, routinely checked for their approximate
validity, and not chosen based on necessary requirements
for classical statistical procedures that have no real scientific merit.

Therefore, a better strategy than focusing on satisfying additivity to
use the F-test for testing Neyman's null, we believe,
is to introduce a Bayesian framework into the problem (\citep
{Rubin1978}). One can obtain a posterior predictive distribution
for the estimand of interest (defined in terms of the potential
outcomes) and evaluate relevant Bayes' rules using the same
criteria that Neyman and others have considered (e.g., consistency,
coverage, Type I error)
(\citep{Rubin1984}). The Fisher
randomization test can be viewed as a type of posterior predictive
check
(\citep{Rubin1984}), and it can be more enlightening
(as the example in Section~\ref{sec:concrete} illustrates) to perform
explicitly the Fisher randomization test for Fisher's
sharp null, rather than using the F-distribution as an approximation
when testing Neyman's null under additivity. When
additivity may not hold, evaluating Bayes' rules motivated by the
particular applied setting of a problem appears to be a
more viable path to the solution of a specific problem than relying on
classical statistical procedures that are imprecise
without applied contexts.

% zodis "Acknowledgments" paliekamas pagal autoriu
\section*{Acknowledgments}
We are grateful to the Executive Editor, an Associate
Editor and a referee for many
valuable comments that improved this paper.
The research of Arman Sabbaghi was
supported by the United States National Science Foundation Graduate Research Fellowship
under Grant No. DGE-1144152.

\begin{supplement}[id=suppA]
\stitle{Supplementary materials for ``Comments on the Neyman--Fisher
Controversy and its Consequences''}
\slink[doi]{10.1214/13-STS454SUPP} %[doi,text={...}] - jei reikia suskaldyti doi
\sdatatype{.pdf}
\sfilename{sts454\_supp.pdf}
\sdescription{The supplementary material contains our reworking of
Neyman's calculations, specifically expectations and
variances of sample averages, and expectations of sums of squares for
RCB and LS designs. These calculations form the basis
of all results presented in this article. The supplementary material
can be accessed via the following link:
\url{http://www.people.fas.harvard.edu/\textasciitilde sabbaghi/sabbaghi\_rubin\_supplement.pdf}.}
\end{supplement}

% imsref loaded by akundreckaite, 2013--12-18 15:49:58
% imsref loaded by akundreckaite, 2013-12-19 08:19:05
%


\begin{thebibliography}{40}
% Style name=ims, version=2.4, label_style=nameyear,
%sorting_style=complex, cfg=None, language=None.

%b1 #&#
%b1 ###
\bibitem[\protect\citeauthoryear{Bartlett}{1947}]{Bartlett1947}
%
\begin{barticle}[mr]
\bauthor{\bsnm{Bartlett},~\bfnm{M.~S.}\binits{M.~S.}}
(\byear{1947}).
\btitle{The use of transformations}.
\bjournal{Biometrics}
\bvolume{3}
\bpages{39--52}.
\bid{issn={0006-341X}, mr={0020763}}
\end{barticle}
%
\bptok{imsref}%
\endbibitem

%b2 #&#
%b2 ###
\bibitem[\protect\citeauthoryear{Box}{1978}]{Box}
%
\begin{bbook}[mr]
\bauthor{\bsnm{Box},~\bfnm{Joan~Fisher}\binits{J.~F.}}
(\byear{1978}).
\btitle{R. {A}. {F}isher: The Life of a Scientist}.
\bpublisher{Wiley},
\blocation{New York}.
\bid{mr={0500579}}
\end{bbook}
%
\bptok{imsref}\vadjust{\goodbreak}%
\endbibitem

%b3 #&#
%b3 ###
\bibitem[\protect\citeauthoryear{Box}{1984}]{Box1984}
%
\begin{barticle}[auto:STB|2013/12/09|07:59:19]
\bauthor{\bsnm{Box},~\bfnm{G.~E.~P.}\binits{G.~E.~P.}}
(\byear{1984}).
\btitle{Discussion of paper by D.{R}. Cox}.
\bjournal{International Statistical Review}
\bvolume{52}
\bpages{26}.
\end{barticle}
%
\bptok{imsref}%
\endbibitem

%b4 #&#
%b4 ###
\bibitem[\protect\citeauthoryear{Box and Cox}{1964}]{BoxCox}
%
\begin{barticle}[mr]
\bauthor{\bsnm{Box},~\bfnm{G.~E.~P.}\binits{G.~E.~P.}} \AND
\bauthor{\bsnm{Cox},~\bfnm{D.~R.}\binits{D.~R.}}
(\byear{1964}).
\btitle{An analysis of transformations ({w}ith discussion)}.
\bjournal{J. R. Stat. Soc. Ser. B Stat. Methodol.}
\bvolume{26}
\bpages{211--252}.
\bid{issn={0035-9246}, mr={0192611}}
\bptnote{check related}%
\end{barticle}
%
\bptok{imsref}%
\endbibitem

%b5 #&#
%b5 ###
\bibitem[\protect\citeauthoryear{Cochran}{1947}]{Cochran1947}
%
\begin{barticle}[mr]
\bauthor{\bsnm{Cochran},~\bfnm{W.~G.}\binits{W.~G.}}
(\byear{1947}).
\btitle{Some consequences when the assumptions for the analysis of
variance are not satisfied}.
\bjournal{Biometrics}
\bvolume{3}
\bpages{22--38}.
\bid{issn={0006-341X}, mr={0020762}}
\end{barticle}
%
\bptok{imsref}%
\endbibitem

%b6 #&#
%b6 ###
\bibitem[\protect\citeauthoryear{Cox}{1958}]{Cox1958}
%
\begin{barticle}[auto:STB|2013/12/09|07:59:19]
\bauthor{\bsnm{Cox},~\bfnm{D.~R.}\binits{D.~R.}}
(\byear{1958}).
\btitle{The interpretation of the effects of non-additivity in the
Latin square}.
\bjournal{Biometrika}
\bvolume{46}
\bpages{69--73}.
\end{barticle}
%
\bptok{imsref}%
\endbibitem

%b7 #&#
%b7 ###
\bibitem[\protect\citeauthoryear{Cox}{1984}]{Cox1984}
%
\begin{barticle}[mr]
\bauthor{\bsnm{Cox},~\bfnm{D.~R.}\binits{D.~R.}}
(\byear{1984}).
\btitle{Interaction}.
\bjournal{Internat. Statist. Rev.}
\bvolume{52}
\bpages{1--31}.
\bid{doi={10.2307/1403235}, issn={0306-7734}, mr={0967201}}
\bptnote{check related}%
\end{barticle}
%
\bptok{imsref}%
\endbibitem

%b8 #&#
%b8 ###
\bibitem[\protect\citeauthoryear{Cox}{1958}]{Cox}
%
\begin{bbook}[mr]
\bauthor{\bsnm{Cox},~\bfnm{D.~R.}\binits{D.~R.}}
(\byear{1958}).
\btitle{Planning of Experiments},
\bedition{1st} ed.
\bpublisher{Wiley},
\blocation{New York}.
\bptnote{check year}%
\end{bbook}
%
\bptok{imsref}%
\endbibitem

%b9 #&#
%b9 ###
\bibitem[\protect\citeauthoryear{Cox}{2012}]{Cox2012}
%
\begin{bincollection}[auto:STB|2013/12/09|07:59:19]
\bauthor{\bsnm{Cox},~\bfnm{D.~R.}\binits{D.~R.}}
(\byear{2012}).
\btitle{Statistical causality: Some historical remarks}.
In \bbooktitle{Causality: Statistical Perspectives and Applications}
(\beditor{\bfnm{C.}\binits{C.}~\bsnm{Berzuini}},
\beditor{\bfnm{P.}\binits{P.}~\bsnm{Dawid}} \AND
\beditor{\bfnm{L.}\binits{L.}~\bsnm{Bernardinelli}}, eds.)
\bpages{1--5}.
\bpublisher{Wiley},
\blocation{New York}.
\end{bincollection}
%
\bptok{imsref}%
\endbibitem




%b11 #&#
%b10 ###
\bibitem[\protect\citeauthoryear{Eisenhart}{1947}]{Eisenhart1947}
%
\begin{barticle}[mr]
\bauthor{\bsnm{Eisenhart},~\bfnm{Churchill}\binits{C.}}
(\byear{1947}).
\btitle{The assumptions underlying the analysis of variance}.
\bjournal{Biometrics}
\bvolume{3}
\bpages{1--21}.
\bid{issn={0006-341X}, mr={0020761}}
\end{barticle}
%
\bptok{imsref}%
\endbibitem

%b12 #&#
%b11 ###
\bibitem[\protect\citeauthoryear{Fienberg and Tanur}{1996}]{FienbergTanur1996}
%
\begin{barticle}[auto:STB|2013/12/09|07:59:19]
\bauthor{\bsnm{Fienberg},~\bfnm{S.~E.}\binits{S.~E.}} \AND
\bauthor{\bsnm{Tanur},~\bfnm{J.~M.}\binits{J.~M.}}
(\byear{1996}).
\btitle{Reconsidering the fundamental contributions of Fisher and
Neyman on experimentation and sampling}.
\bjournal{International Statistical Review}
\bvolume{64}
\bpages{237--253}.
\end{barticle}
%
\bptok{imsref}%
\endbibitem

%b13 #&#
%b12 ###
\bibitem[\protect\citeauthoryear{Fisher}{1935}]{Fisher1935}
%
\begin{barticle}[auto:STB|2013/12/09|07:59:19]
\bauthor{\bsnm{Fisher},~\bfnm{R.~A.}\binits{R.~A.}}
(\byear{1935}).
\btitle{Comment on ``Statistical problems in agricultural
experimentation (with discussion).''}
\bjournal{Suppl. J. Roy. Statist. Soc. Ser. B}
\bvolume{2}
\bpages{154--157, 173}.
\end{barticle}
%
\bptok{imsref}%
\endbibitem

%b14 #&#
%b13 ###
\bibitem[\protect\citeauthoryear{Fisher}{1971}]{Fisher}
%
\begin{bmisc}[auto:STB|2013/12/09|07:59:19]
\bauthor{\bsnm{Fisher},~\bfnm{R.~A.}\binits{R.~A.}}
(\byear{1971}).
\bhowpublished{\textit{The Design of Experiments},
9th ed. Hafner Publishing Company, New York}.
\end{bmisc}
%
\bptok{imsref}%
% NOT OUTPUTED:
% sortkey = Fisher(1971
\endbibitem

%b15 #&#
%b14 ###
\bibitem[\protect\citeauthoryear{Gourlay}{1955a}]{Gourlay1955a}
%
\begin{barticle}[auto:STB|2013/12/09|07:59:19]
\bauthor{\bsnm{Gourlay},~\bfnm{N.}\binits{N.}}
(\byear{1955}a).
\btitle{F-test bias for experimental designs in educational research}.
\bjournal{Psychometrika}
\bvolume{20}
\bpages{227--258}.
\end{barticle}
%
\bptok{imsref}%
\endbibitem

%b16 #&#
%b15 ###
\bibitem[\protect\citeauthoryear{Gourlay}{1955b}]{Gourlay1955b}
%
\begin{barticle}[auto:STB|2013/12/09|07:59:19]
\bauthor{\bsnm{Gourlay},~\bfnm{N.}\binits{N.}}
(\byear{1955}b).
\btitle{F-test bias for experimental designs of the latin square type}.
\bjournal{Psychometrika}
\bvolume{20}
\bpages{273--287}.
\end{barticle}
%
\bptok{imsref}%
\endbibitem

%b17 #&#
%b16 ###
\bibitem[\protect\citeauthoryear{Hinkelmann and Kempthorne}{2008}]{Kempthorne}
%
\begin{bbook}[mr]
\bauthor{\bsnm{Hinkelmann},~\bfnm{Klaus}\binits{K.}} \AND
\bauthor{\bsnm{Kempthorne},~\bfnm{Oscar}\binits{O.}}
(\byear{2008}).
\btitle{Design and Analysis of Experiments. {V}ol. 1: Introduction to
Experimental Design},
\bedition{2nd} ed.
\bpublisher{Wiley},
\blocation{Hoboken, NJ}.
\bid{mr={2363107}}
\end{bbook}
%
\bptok{imsref}%
\endbibitem

%b18 #&#
%b17 ###
\bibitem[\protect\citeauthoryear{Kempthorne}{1952}]{Kempthorne1952}
%
\begin{bbook}[mr]
\bauthor{\bsnm{Kempthorne},~\bfnm{Oscar}\binits{O.}}
(\byear{1952}).
\btitle{The Design and Analysis of Experiments}.
\bpublisher{Wiley},
\blocation{New York}.
\bid{mr={0045368}}
\end{bbook}
%
\bptok{imsref}%
\endbibitem

%b19 #&#
%b18 ###
\bibitem[\protect\citeauthoryear{Kempthorne}{1955}]{Kempthorne1955}
%
\begin{barticle}[mr]
\bauthor{\bsnm{Kempthorne},~\bfnm{Oscar}\binits{O.}}
(\byear{1955}).
\btitle{The randomization theory of experimental inference}.
\bjournal{J. Amer. Statist. Assoc.}
\bvolume{50}
\bpages{946--967}.
\bid{issn={0162-1459}, mr={0071696}}
\end{barticle}
%
\bptok{imsref}%
\endbibitem

%b20 #&#
%b19 ###
\bibitem[\protect\citeauthoryear{Lehmann}{2011}]{Lehmann}
%
\begin{bbook}[mr]
\bauthor{\bsnm{Lehmann},~\bfnm{Erich~L.}\binits{E.~L.}}
(\byear{2011}).
\btitle{Fisher, {N}eyman, and the Creation of Classical Statistics}.
\bpublisher{Springer},
\blocation{New York}.
\bid{doi={10.1007/978-1-4419-9500-1}, mr={2798202}}
\end{bbook}
%
\bptok{imsref}%
\endbibitem

%b21 #&#
%b20 ###
\bibitem[\protect\citeauthoryear{Mandel}{1961}]{Mandel}
%
\begin{barticle}[mr]
\bauthor{\bsnm{Mandel},~\bfnm{John}\binits{J.}}
(\byear{1961}).
\btitle{Non-additivity in two-way analysis of variance}.
\bjournal{J. Amer. Statist. Assoc.}
\bvolume{56}
\bpages{878--888}.
\bid{issn={0162-1459}, mr={0131934}}
\end{barticle}
%
\bptok{imsref}%
\endbibitem

%b22 #&#
%b21 ###
\bibitem[\protect\citeauthoryear{Neyman}{1935}]{Neyman1935}
%
\begin{barticle}[auto:STB|2013/12/09|07:59:19]
\bauthor{\bsnm{Neyman},~\bfnm{J.}\binits{J.}}
(\byear{1935}).
\btitle{Statistical problems in agricultural experimentation (with discussion)}.
\bjournal{Suppl. J. Roy. Statist. Soc. Ser. B}
\bvolume{2}
\bpages{107--180}.
\end{barticle}
%
\bptok{imsref}%
\endbibitem

%b23 #&#
%b22 ###
\bibitem[\protect\citeauthoryear{Neyman}{1976}]{Neyman1976}
%
\begin{bincollection}[mr]
\bauthor{\bsnm{Neyman},~\bfnm{J.}\binits{J.}}
(\byear{1976}).
\btitle{Emergence of mathematical statistics}.
In \bbooktitle{On the History of Statistics and Probability:
Proceedings of a Symposium on the American Mathematical Heritage, to
Celebrate the Bicentennial of the United States of America, Held at
Southern Methodist University, May 27--29, 1974}
(\beditor{\bfnm{D.~B.}\binits{D.~B.}~\bsnm{Owen}},
\beditor{\bfnm{W.~G.}\binits{W.~G.}~\bsnm{Cochran}},
\beditor{\bfnm{H.~O.}\binits{H.~O.}~\bsnm{Hartley}} \AND
\beditor{\bfnm{J.}\binits{J.}~\bsnm{Neyman}}, eds.)
\bpages{149--185}.
\bpublisher{Dekker},
\blocation{New York}.
\end{bincollection}
%
\bptok{imsref}%
\endbibitem



%b24 #&#
%b23 ###
\bibitem[\protect\citeauthoryear{Pitman}{1938}]{Pitman}
%
\begin{barticle}[auto:STB|2013/12/09|07:59:19]
\bauthor{\bsnm{Pitman},~\bfnm{E.}\binits{E.}}
(\byear{1938}).
\btitle{Significance tests which may be applied to samples from any
populations: III. The Analysis of Variance Test}.
\bjournal{Biometrika}
\bvolume{29}
\bpages{322--335}.
\end{barticle}
%
\bptok{imsref}%
\endbibitem

%b25 #&#
%b24 ###
\bibitem[\protect\citeauthoryear{Reid}{1982}]{Reid}
%
\begin{bbook}[mr]
\bauthor{\bsnm{Reid},~\bfnm{Constance}\binits{C.}}
(\byear{1982}).
\btitle{Neyman---from Life}.
\bpublisher{Springer},
\blocation{New York}.
\bid{doi={10.1007/978-1-4612-5754-7}, mr={0680939}}
\end{bbook}
%
\bptok{imsref}%
\endbibitem

%b26 #&#
%b25 ###
\bibitem[\protect\citeauthoryear{Rojas}{1973}]{Rojas}
%
\begin{barticle}[auto:STB|2013/12/09|07:59:19]
\bauthor{\bsnm{Rojas},~\bfnm{B.}\binits{B.}}
(\byear{1973}).
\btitle{On Tukey's test of additivity}.
\bjournal{Biometrics}
\bvolume{29}
\bpages{45--52}.
\end{barticle}
%
\bptok{imsref}%
\endbibitem

%b27 #&#
%b26 ###
\bibitem[\protect\citeauthoryear{Rubin}{1978}]{Rubin1978}
%
\begin{barticle}[mr]
\bauthor{\bsnm{Rubin},~\bfnm{Donald~B.}\binits{D.~B.}}
(\byear{1978}).
\btitle{Bayesian inference for causal effects: The role of randomization}.
\bjournal{Ann. Statist.}
\bvolume{6}
\bpages{34--58}.
\bid{issn={0090-5364}, mr={0472152}}
\end{barticle}
%
\bptok{imsref}%
\endbibitem

%b28 #&#
%b27 ###
\bibitem[\protect\citeauthoryear{Rubin}{1984}]{Rubin1984}
%
\begin{barticle}[mr]
\bauthor{\bsnm{Rubin},~\bfnm{Donald~B.}\binits{D.~B.}}
(\byear{1984}).
\btitle{Bayesianly justifiable and relevant frequency calculations for
the applied statistician}.
\bjournal{Ann. Statist.}
\bvolume{12}
\bpages{1151--1172}.
\bid{doi={10.1214/aos/1176346785}, issn={0090-5364}, mr={0760681}}
\end{barticle}
%
\bptok{imsref}%
\endbibitem

%b29 #&#
%b28 ###
\bibitem[\protect\citeauthoryear{Rubin}{1990}]{Rubin1990}
%
\begin{barticle}[mr]
\bauthor{\bsnm{Rubin},~\bfnm{Donald~B.}\binits{D.~B.}}
(\byear{1990}).
\btitle{Comment on {J}. {N}eyman and causal inference in experiments
and observational studies: ``{O}n the application of
probability theory to agricultural experiments.
{E}ssay on principles. {S}ection 9''
[\textit{{A}nn. {A}gric. {S}ci.} \textbf{10} (1923) 1--51]}.
\bjournal{Statist. Sci.}
\bvolume{5}
\bpages{472--480}.
\bid{issn={0883-4237}, mr={1092987}}
\end{barticle}
%
\bptok{imsref}%
\endbibitem

%b30 #&#
%b29 ###
\bibitem[\protect\citeauthoryear{Rubin}{2005}]{Rubin2005}
%
\begin{barticle}[mr]
\bauthor{\bsnm{Rubin},~\bfnm{Donald~B.}\binits{D.~B.}}
(\byear{2005}).
\btitle{Causal inference using potential outcomes: Design, modeling, decisions}.
\bjournal{J. Amer. Statist. Assoc.}
\bvolume{100}
\bpages{322--331}.
\bid{doi={10.1198/016214504000001880}, issn={0162-1459}, mr={2166071}}
\end{barticle}
%
\bptok{imsref}%
\endbibitem

%b31 #&#
%b30 ###
\bibitem[\protect\citeauthoryear{Sabbaghi and Rubin}{2014}]{SabbaghiRubin}
%
\begin{bmisc}[auto:STB|2013/12/09|07:59:19]
\bauthor{\bsnm{Sabbaghi},~\bfnm{A.}\binits{A.}} \AND
\bauthor{\bsnm{Rubin},~\bfnm{D.~B.}\binits{D.~B.}}
(\byear{2014})
\bhowpublished{Supplement to ``Comments on the Neyman--{F}isher
controversy and its consequences.'' DOI:\doiurl{10.1214/13-STS454SUPP}.}
\end{bmisc}
%
\bptok{imsref}%
% NOT OUTPUTED:
% sortkey = Sabbaghi(2014
\endbibitem


%b32 #&#
%b31 ###
\bibitem[\protect\citeauthoryear{Splawa-Neyman}{1990}]{Neyman1923}
%
\begin{barticle}[mr]
\bauthor{\bsnm{Splawa-Neyman},~\bfnm{Jerzy}\binits{J.}}
(\byear{1990}).
\btitle{On the application of probability theory to agricultural
experiments. {E}ssay on principles. {S}ection~9}.
\bjournal{Statist. Sci.}
\bvolume{5}
\bpages{465--472}.
\bid{issn={0883-4237}, mr={1092986}}
\end{barticle}
%
\bptok{imsref}%
\endbibitem

%b33 #&#
%b32 ###
\bibitem[\protect\citeauthoryear{Sukhatme}{1935}]{Sukhatme}
%
\begin{barticle}[auto:STB|2013/12/09|07:59:19]
\bauthor{\bsnm{Sukhatme},~\bfnm{P.}\binits{P.}}
(\byear{1935}).
\btitle{Comment on ``Statistical problems in agricultural
experimentation (with discussion).''}
\bjournal{Suppl. J. Roy. Statist. Soc. Ser. B}
\bvolume{2}
\bpages{166--169}.
\end{barticle}
%
\bptok{imsref}%
\endbibitem

%b34 #&#
%b33 ###
\bibitem[\protect\citeauthoryear{Tukey}{1949}]{Tukey}
%
\begin{barticle}[auto:STB|2013/12/09|07:59:19]
\bauthor{\bsnm{Tukey},~\bfnm{J.}\binits{J.}}
(\byear{1949}).
\btitle{One degree of freedom for nonadditivity}.
\bjournal{Biometrics}
\bvolume{5}
\bpages{232--242}.
\end{barticle}
%
\bptok{imsref}%
\endbibitem

%b10 #&#
%b34 ###
\bibitem[\protect\citeauthoryear{Tukey}{1955}]{Tukey1955}
%
\begin{barticle}[auto]
\bauthor{\bsnm{Tukey},~\bfnm{J.}\binits{J.}}
(\byear{1955}).
\btitle{Query 113}.
\bjournal{Biometrics}
\bvolume{11}
\bpages{111--113}.
\end{barticle}
%
\bptok{imsref}%
\endbibitem

%b35 #&#
%b35 ###
\bibitem[\protect\citeauthoryear{Welch}{1937}]{Welch}
%
\begin{barticle}[auto:STB|2013/12/09|07:59:19]
\bauthor{\bsnm{Welch},~\bfnm{B.}\binits{B.}}
(\byear{1937}).
\btitle{On the z-test in randomized blocks and latin squares}.
\bjournal{Biometrika}
\bvolume{29}
\bpages{21--52}.
\end{barticle}
%
\bptok{imsref}%
\endbibitem

%b36 #&#
%b36 ###
\bibitem[\protect\citeauthoryear{Wilk}{1955}]{Wilk1955}
%
\begin{barticle}[mr]
\bauthor{\bsnm{Wilk},~\bfnm{M.~B.}\binits{M.~B.}}
(\byear{1955}).
\btitle{The randomization analysis of a generalized randomized block design}.
\bjournal{Biometrika}
\bvolume{42}
\bpages{70--79}.
\bid{issn={0006-3444}, mr={0068800}}
\end{barticle}
%
\bptok{imsref}%
\endbibitem

%b37 #&#
%b37 ###
\bibitem[\protect\citeauthoryear{Wilk and Kempthorne}{1957}]{WilkKempthorne}
%
\begin{barticle}[mr]
\bauthor{\bsnm{Wilk},~\bfnm{M.~B.}\binits{M.~B.}} \AND
\bauthor{\bsnm{Kempthorne},~\bfnm{Oscar}\binits{O.}}
(\byear{1957}).
\btitle{Non-additivities in a {L}atin square design}.
\bjournal{J. Amer. Statist. Assoc.}
\bvolume{52}
\bpages{218--236}.
\bid{issn={0162-1459}, mr={0088137}}
\end{barticle}
%
\bptok{imsref}%
\endbibitem

%b38 #&#
%b38 ###
\bibitem[\protect\citeauthoryear{Wu and Hamada}{2009}]{wuhamada}
%
\begin{bbook}[mr]
\bauthor{\bsnm{Wu},~\bfnm{C.~F.~Jeff}\binits{C.~F.~J.}} \AND
\bauthor{\bsnm{Hamada},~\bfnm{Michael~S.}\binits{M.~S.}}
(\byear{2009}).
\btitle{Experiments: Planning, Analysis, and Optimization},
\bedition{2nd} ed.
\bpublisher{Wiley},
\blocation{Hoboken, NJ}.
\bid{mr={2583259}}
\end{bbook}
%
\bptok{imsref}%
\endbibitem

%b39 #&#
%b39 ###
\bibitem[\protect\citeauthoryear{Yates}{1935}]{Yates1935}
%
\begin{barticle}[auto:STB|2013/12/09|07:59:19]
\bauthor{\bsnm{Yates},~\bfnm{F.}\binits{F.}}
(\byear{1935}).
\btitle{Complex experiments}.
\bjournal{J. R. Stat. Soc. Ser. B Stat. Methodol.}
\bvolume{2}
\bpages{181--247}.
\end{barticle}
%
\bptok{imsref}%
\endbibitem

%b40 #&#
%b40 ###
\bibitem[\protect\citeauthoryear{Yates}{1939}]{Yates}
%
\begin{barticle}[auto:STB|2013/12/09|07:59:19]
\bauthor{\bsnm{Yates},~\bfnm{F.}\binits{F.}}
(\byear{1939}).
\btitle{The comparative advantages of systematic and randomized
arrangements in the design of agricultural and biological experiments}.
\bjournal{Biometrika}
\bvolume{30}
\bpages{440--466}.
\end{barticle}
%
\bptok{imsref}%
\endbibitem

\end{thebibliography}
\end{document}